\documentclass[amsmath,amssymb,aps,pre,showpacs,twocolumn,floatfix,10pt]{revtex4-1}
\pdfoutput=1
\usepackage{graphicx}
\usepackage{color}
\begin{document}
\title{Pattern formation in fast-growing sandpiles}
\author{Tridib Sadhu$^{1,2}$}
\author{Deepak Dhar$^1$}

\affiliation{$^1$Department of Theoretical Physics \\ Tata Institute of
Fundamental Research \\ Mumbai 400005, India.\\
~\\
$^2$Department Physics of Complex Systems\\ Weizmann Institute
of Science\\ Rehovot 76100, Israel.}

\begin{abstract}
We study the patterns formed by adding  $N$ sand-grains at a single
site on an initial periodic background in the
Abelian sandpile models, and relaxing the configuration. When the heights at
all sites in the initial background are low enough, one gets patterns
showing proportionate growth, with the diameter of the pattern formed
growing  as $N^{1/d}$ for large $N$, in $d$-dimensions. On the other
hand, if sites with
maximum stable height in the starting configuration form an infinite
cluster, we get  avalanches that do not stop. In this paper, we
describe our unexpected finding of an interesting class of
backgrounds in two dimensions,  that  show  an  intermediate behavior:
For any $N$, the avalanches are finite, but  the diameter of the
pattern increases   as  $N^{\alpha}$, for large $N$, with $1/2 <
\alpha \leq 1$.  Different values of $\alpha$ can be realized  on
different backgrounds, and the patterns still show proportionate
growth. The non-compact nature of growth simplifies their
analysis significantly. We characterize the asymptotic pattern exactly for one
illustrative example with $\alpha=1$.
\end{abstract}
\pacs{89.75.Kd, 45.70.Cc, 05.65.+b}
\maketitle

\section{Introduction}\label{sec:intro}
\begin{figure}
\includegraphics[width=8cm,angle=0]{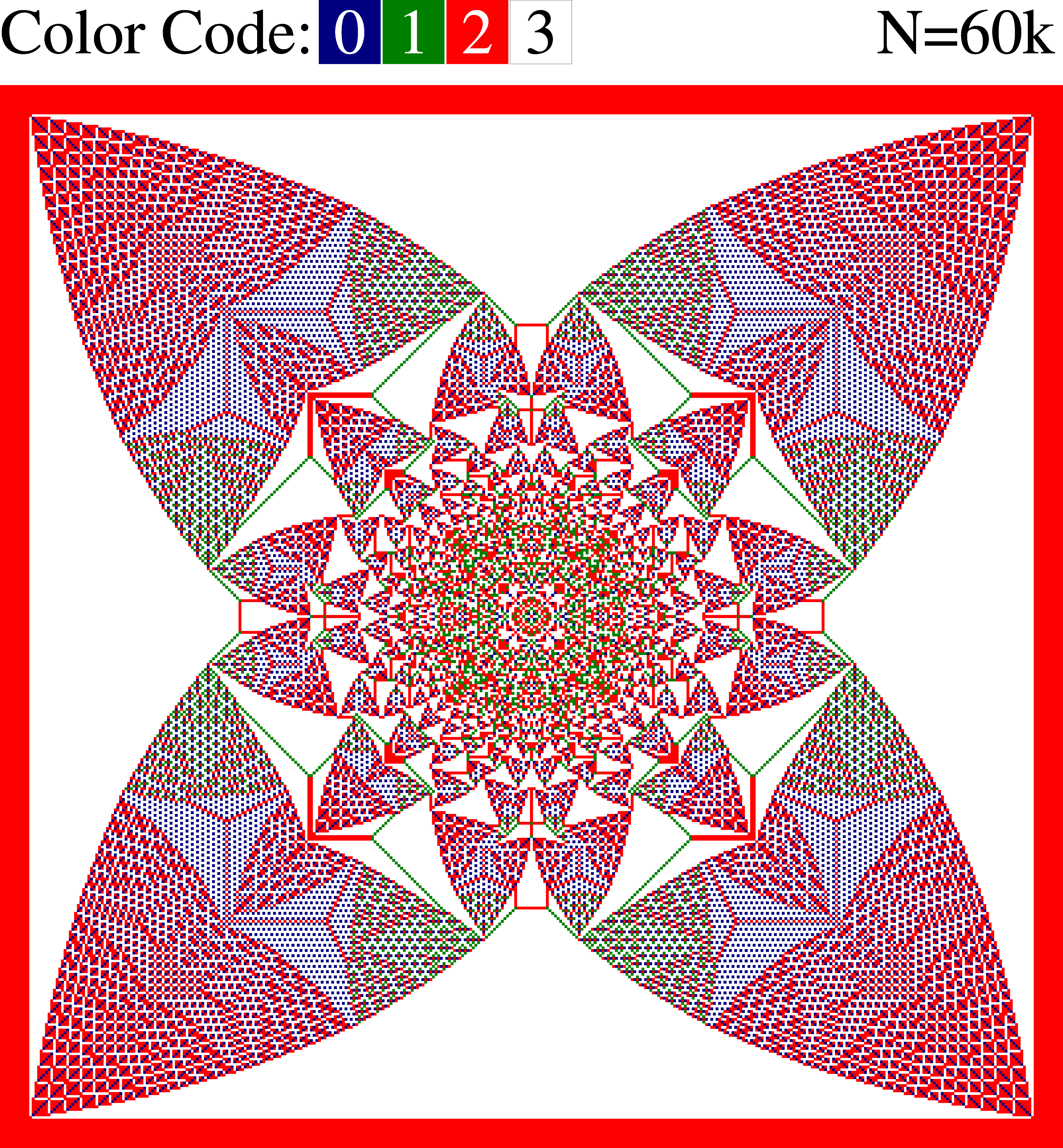}\\
~\\
\includegraphics[width=8cm,angle=0]{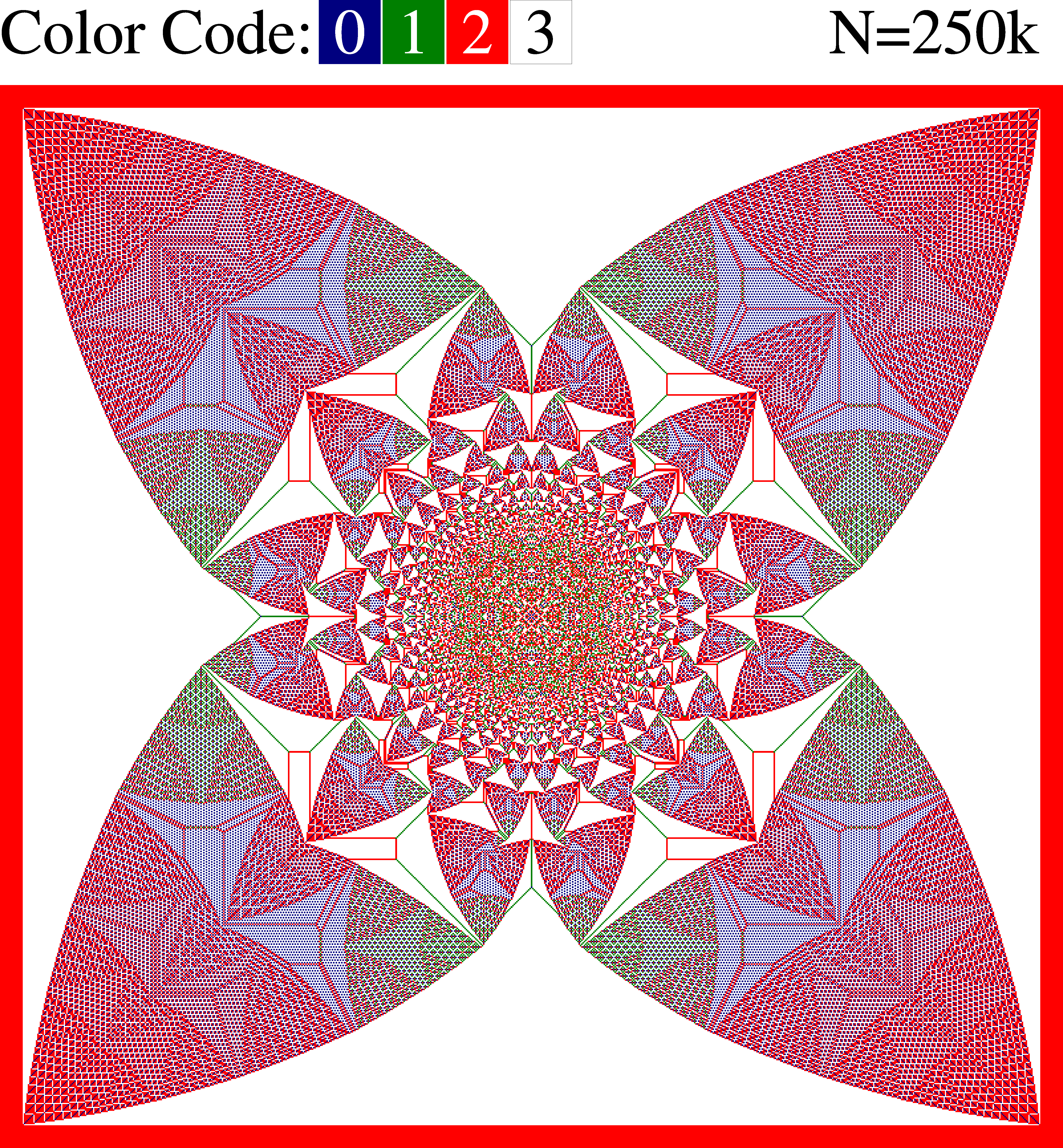}
\caption{(Color online.) The stable configurations for the Abelian sandpile model on a
square lattice, obtained by adding $N$ grains at the origin. In the
initial configuration all heights are $2$. For comparison the size of
the first figure has been enlarged by a factor $2$. (Details can be
seen in the electronic version using zoom in).}
\label{fig:btw} 
\end{figure}

In the last two decades a large amount of study has been  devoted to
understanding various models of self-organized criticality, in
particular, the Abelian Sandpile Model (ASM) (see
\cite{dhar2006,redigleshouches} for reviews). These have  mainly dealt with the critical exponents of
avalanches produced in sandpiles driven slowly in their critical
steady state. But the ASM has other interesting properties, not
directly related to its critical exponents. In particular, one
sees  very interesting and beautiful spatial patterns when many sand
grains are added at a \textit{single} point on an initially periodic  background,
and we relax the configuration using the ASM toppling rules. One such
pattern on a square lattice is shown in Fig. \ref{fig:btw}.

Our interest in these patterns comes from several reasons. Firstly,
these are analytically tractable examples of complex patterns that are
obtained from simple deterministic evolution rules. Of course, there
are many known examples of complex patterns obtained this way (\textit{e.g.} Conway's game of life
\cite{earlierone}). But a detailed characterization of such patterns
is usually not easy. The sandpile  patterns studied here are
special, as they are nontrivial, but of intermediate complexity, and
are analytically tractable.

Secondly, growing sandpiles studied here are {\it qualitatively different} from
the growth models that have been studied in physics literature
earlier, such as the Eden model, the diffusion limited aggregation, or
the surface deposition \cite{eden,dla,barabasi}. These are the simplest models of
\textit{proportionate growth}, a well-known feature of
biological growth in animals, where different parts of a growing
animal grow at roughly the same rate, keeping their shape almost the
same.  In the models of growth studied earlier, growth
is confined to some active outer region. The inner structures, once
formed are frozen, and do not evolve further in time. This is not the
case for the patterns studied in this paper. Figure \ref{fig:btw} shows two patterns produced
on the same background but with different values of the number $N$ of
grains added. As $N$ increases, the pattern grows in size, but we see that while some new details emerge near the center, the relative proportions of different parts in the outer region of the pattern
remains unchanged. As the pattern grows, different features, not only
grow in size, but also are moved in space with time.

The third motivation for our study is some intriguing connection of
these to the mathematics of discrete analytic functions. We have not
explored this much, but will discuss briefly in an appendix.

There have been several earlier studies of the spatial patterns in sandpile
models. First of them was by Liu \textit{et.al.} \cite{liu}. The
asymptotic shape of the boundaries of the patterns produced in
centrally seeded sandpile model on different periodic backgrounds was
discussed in \cite{dhar99}. Borgne \textit{et.al.} \cite{borgne}
obtained bounds on the rate of growth of these boundaries, and later
these bounds were improved by Fey \textit{et.al.} \cite{redig} and
Levine \textit{et.al.} \cite{lionel2}. A detailed analysis of
different periodic structures found in the patterns were first carried
out by Ostojic \cite{ostojic} who also first noted the exact quadratic nature
of the toppling function within a patch. Wilson \textit{et.al.} \cite{wilson} have developed a
very efficient algorithm to generate patterns for a large numbers of
particles added, which allows them to generate pictures of patterns
with $N$ up to $2^{26}$.

Other special configurations in the Abelian sandpile models, like the
identity \cite{borgne,identity,caracciolo} or the stable state
produced from special unstable states, also show complex internal
self-similar structures \cite{liu}, which share common features with
the patterns studied here. In particular, the identity configuration
on the F-lattice has recently been shown to have similar spatial
structures \cite{caracciolo}.

There are other models, which are related to the Abelian sandpile
model, \textit{e.g.}, the Internal Diffusion-Limited Aggregation
(IDLA) \cite{lawler},
Eulerian walkers (also called the rotor-router model)
\cite{abhishek,lionel1,propprotor}, and the
infinitely-divisible sandpile \cite{lionel2}, which also show similar structure. For the IDLA, Gravner and Quastel showed that
the asymptotic shape of the growth pattern is related to the classical
Stefan problem in hydrodynamics, and determined the exact radius of
the pattern with a single point source \cite{gravner}. Levine and
Peres have studied patterns with multiple sources in these models, and
proved the existence of a limit shape\cite{lionel3}. Limiting
shapes for the non-Abelian sandpile has recently been studied by Fey.
\textit{et.al.} \cite{nonab}.

\begin{figure}
\includegraphics[width=8cm,clip]{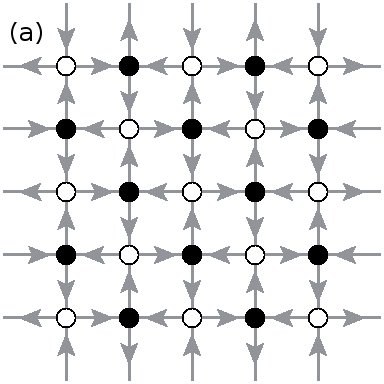}\\
~\\
\includegraphics[width=8cm,angle=0]{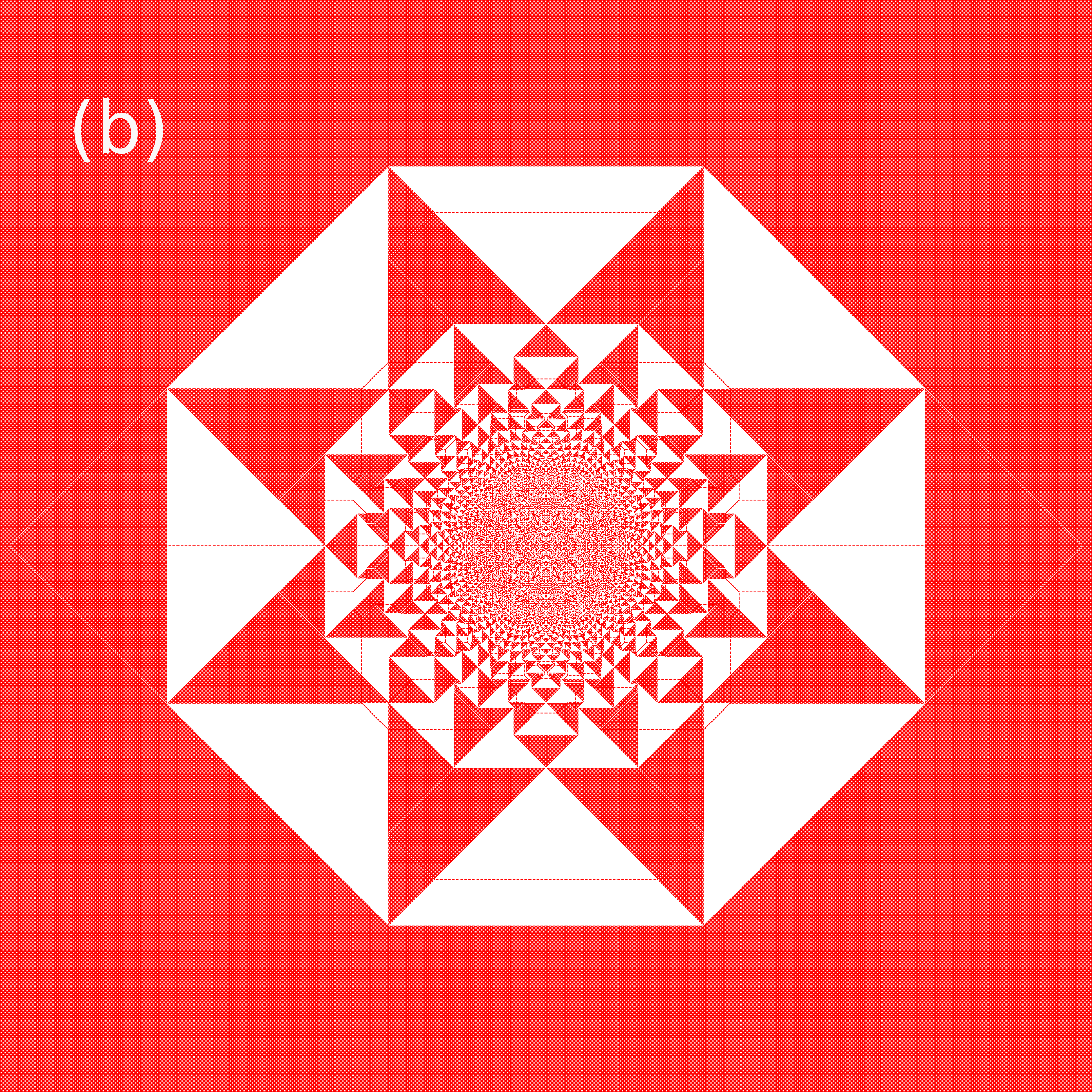}
\caption{(Color online.) (a) A F-lattice with directed bonds, and a chequerboard
distribution of grains on it. Unfilled circles
denote height $z=1$ and filled ones $z=0$. (b) The stable
configuration  obtained by adding $10^5$ grains at the origin.
Color code: red =0, white = 1. The apparent orange regions in the
picture represent the patches with checkerboard configuration.
(Details can be seen in the electronic version using zoom in.)}
\label{fig:flatone}
\end{figure}

The standard square lattice produces complicated patterns and it has
not been possible to characterize them fully, so far. In an earlier
paper \cite{myepl}, we considered the pattern produced on a
F-lattice (Fig. \ref{fig:flatone}(a)), and determined exactly the sizes
of different patches in the asymptotic pattern. The pattern produced
by adding grains at one site on a
background with a periodic chequerboard pattern of
alternate sites with heights $1$ and $0$, is shown in the Fig.
\ref{fig:flatone}(b). In paper \cite{myjsp}, we studied the patterns when
sink sites are present, or when addition is made at more
than one site. In paper \cite{myjsm}, we have studied the effect of
noise on such patterns.

If the average initial height in a background is high, one gets infinite
avalanches, with the diameter of the pattern becoming infinite for
finite number of particles added. Such backgrounds have been termed as
`explosive'. In other cases, the diameter of the pattern is finite for
any finite $N$, and increases as $N^{1/d}$ in $d$-dimensions. We call
such a growth as compact growth. All the patterns studied in
\cite{myepl,myjsp,myjsm} showed compact proportionate growth. In this paper, we
describe a remarkable class of patterns where the diameter remains
finite for any finite $N$, but grows as $N^{\alpha}$, with $1/2 <
\alpha \leq 1$. We call such growth as non-compact proportionate growth.
Characterization of these patterns, as will be shown, is simpler
than in the compact case.

\begin{figure}
\begin{center}
\includegraphics[width=8cm,angle=0]{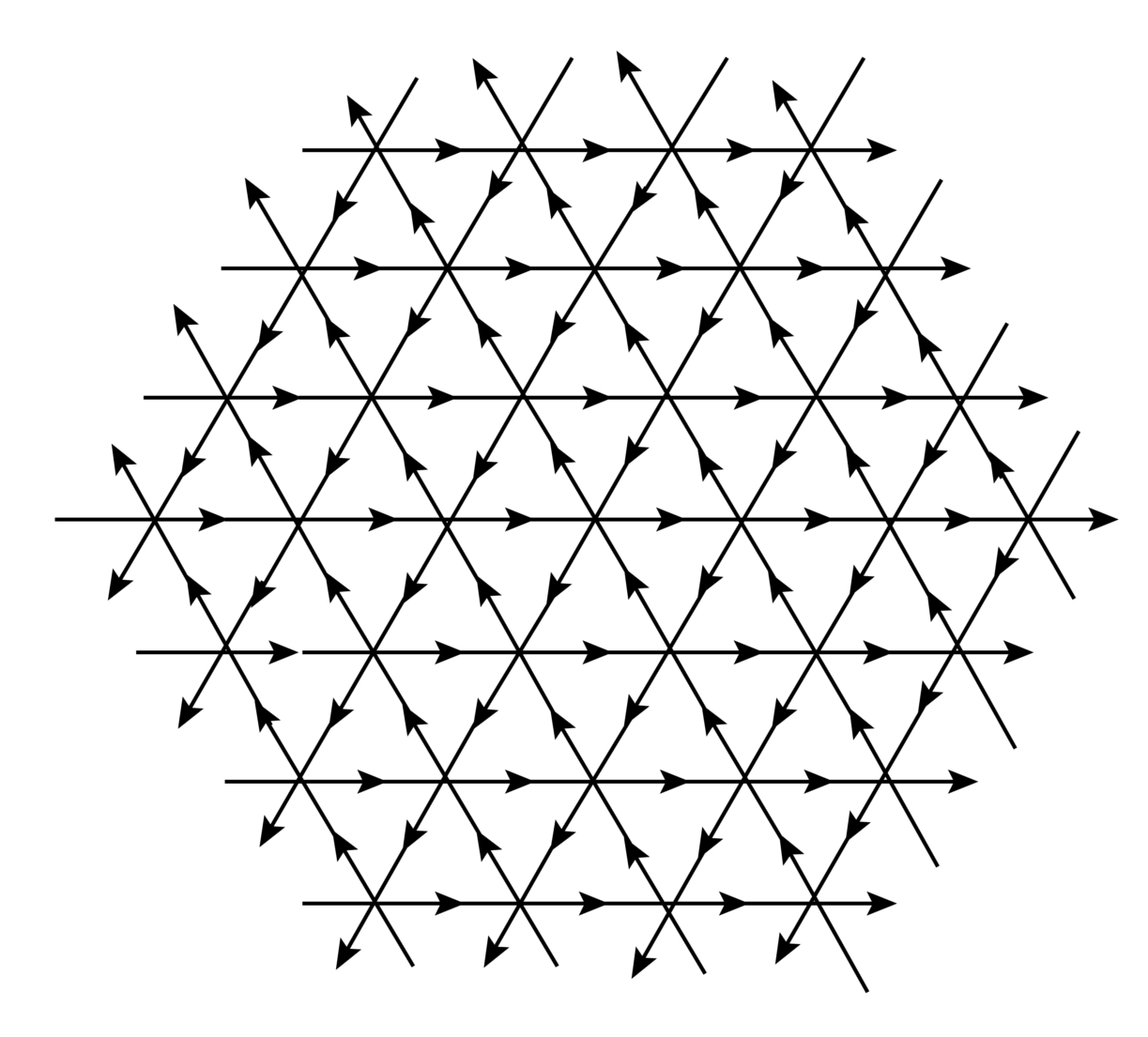}
\caption{A directed triangular lattice.}
\label{fig:trilattice}
\end{center}
\end{figure}

We found two classes of backgrounds, both infinite, on a directed
triangular lattice (see Fig. \ref{fig:trilattice}), for which the
growth is proportionate, with the growth exponent $\alpha=1$. Our
numerical study shows, but we have no formal proof, that  different
backgrounds belonging to  the same class produce the same asymptotic
pattern. In addition, we found infinitely many backgrounds on the
F-lattice which produce patterns with proportionate non-compact
growth. However, in these cases the growth exponent $\alpha$ takes a
different value, with $1/2<\alpha<1$ for each member.

There are some earlier works on the growth rate of the sandpile
patterns. Different non-explosive backgrounds for a DASM were studied in
\cite{redig,lionel2,explosion}. However, in all these examples, studied so far,
the growth of the patterns is compact. For a DASM on a square lattice,
it was shown \cite{explosion}, that the pattern produced on a background of
constant height $z\le z_{c}-2$, is always enclosed inside a square
whose width grows as $\sqrt{N}$.

We also discuss the exact characterization of the pattern shown in
Fig. \ref{fig:hexpicl1}, one of the two  asymptotic patterns we have found
with $\alpha =1$. This is described, as in the earlier studied case of
compact growth \cite{myepl},
in terms of the scaled toppling function. However, the analysis of
non-compact patterns is actually {\it simpler}. Clearly, for
$\alpha > 1/d$, the mean
excess density of particles in the toppled region is zero, for the
asymptotic non-compact growth patterns. The patterns are made of large
regions where the heights are periodic, and we call them as patches. We find that, inside each patch, the mean density is exactly the same as
in the background, and the excess grains are concentrated along the
patch boundaries. There are also some boundaries where excess grains
density is negative. We show that this leads to the scaled toppling
function being a piece-wise linear function of the rescaled coordinates. Thus, in each
patch, the potential function is specified by only $3$ coefficients.
In contrast, for the compact patterns, the scaled toppling function
is a quadratic function of the coordinates in each patch \cite{myepl},
and one has to determine $6$ coefficients for each patch, to
determine the function fully.

We are able to reduce the problem of determining the asymptotic
pattern  in  Fig. \ref{fig:hexpicl1} to that of finding the
lattice Green's function on a hexagonal lattice. This is known to be
expressible as integrals that can be evaluated  in closed form
\cite{atkinson}, and this leads to a full solution of the problem.
For the compact growth patterns studied earlier on F-lattice \cite{myepl,myjsp},
we define a discrete analytic function $F_{p}(z)$, which is the
discrete analogue of the analytic function $z^{p}$, for any rational
value of $p$, and show that the patterns can be characterized in terms of this function.

The plan of this paper is as follows: In section \ref{sec:2}, we discuss how
different periodic background configurations give rise to different
rates of growth. In section \ref{sec:3}, we describe two classes of periodic
background configurations on a directed triangular lattice that give
rise to patterns which show proportionate growth with $\alpha =1$. In
section \ref{sec:4} we argue that, for any pattern with non-compact proportionate
growth the rescaled toppling function is piece-wise linear.
In section \ref{sec:5}, we discuss exact characterization of the
simplest of the patterns with $\alpha=1$. Patterns on the F-lattice, with $ 1/2 < \alpha \leq 1$
are discussed in section \ref{sec:6}.  Section \ref{sec:7} contains some concluding
remarks and a discussion about connection to tropical polynomials. An
appendix describes how the characterization
of these patterns involves the theory of discrete analytic functions
defined on many-sheeted discretized Riemann surfaces.

\section{The compact and non-compact growth patterns\label{sec:2}}
\begin{figure}
\begin{center}
\includegraphics[width=7.0cm]{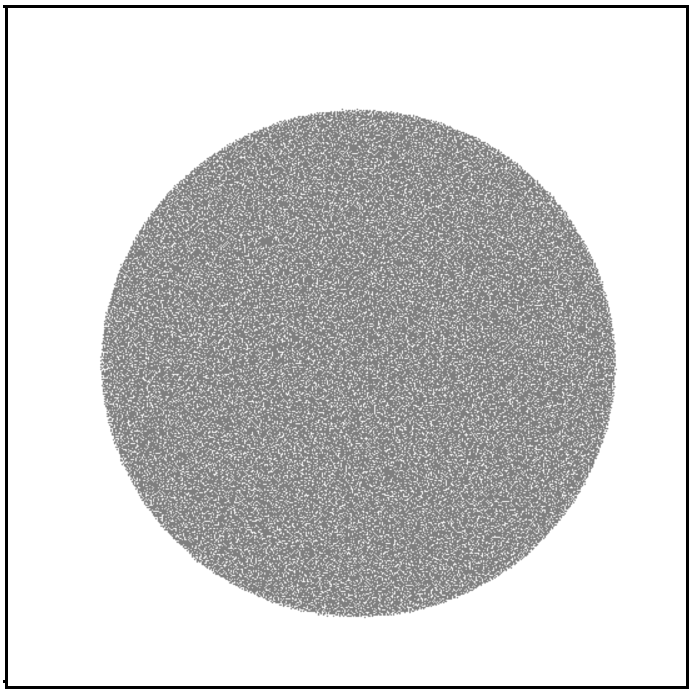}
\caption{The pattern produced by adding $N=10^5$ grains at a single site on a
stochastic ASM defined on an infinite square lattice, and relaxing the
configuration. The initial
configuration has all sites empty. The
threshold height $z_{c}=2$, and on toppling two grains are transfered
either to the vertical or horizontal neighbors, with equal
probability. Color code: White=0, and Black=1.}
\label{fig:mannapattern}
\end{center}
\end{figure}

The simplest growing patterns are found in the  Manna-type sandpile
models with stochastic toppling rules \cite{sasm}. In these models, when the
density of particles $\rho_{o}$ in the background is small,
the avalanches are always finite. In the relaxed configuration, the
toppled sites form a nearly circular region (see Fig.
\ref{fig:mannapattern}). The asymptotic pattern seems to be perfectly
circular disc of uniform density, with an average density $\rho^\star$
inside the circle and $\rho_{o}$ outside. The value of $\rho^\star$ is
independent of the background density $\rho_{o}$, and is equal to the
unique steady state density of the corresponding self organized
critical model with random sites of addition, and dissipation at the
boundary \cite{sasm}. The region inside the circle forgets about the
initial height configuration, and is in the self-organized critical
state. The boundary of the affected region is thin with a sharp
transition of density from $\rho^\star$ to $\rho_{o}$. Then considering
that, for large $N$, all the added grains are confined inside the
circular region of diameter $2\Lambda$, we get
\begin{equation}
N=\left( \rho^\star-\rho_{o} \right)\pi \Lambda^{2}+\rm{Lower~ order~
in~}\Lambda.
\label{eq1}
\end{equation}
Thus the pattern has a compact growth.

For densities $\rho_{o}$ close to, but below  $\rho^\star$, sometimes a
single particle addition can lead to very large increase in the size of the toppled region. However,
probability of such large jumps decreases exponentially with size, and
for any finite $N$, with probability $1$, avalanches remain finite. As
long as $\rho_{o}$ is less than $\rho^\star$, the system relaxes, forming a
pattern whose diameter grows as $\sqrt{N}$. Adding a single grain on a
background of super-critical density ($\rho > \rho^\star$) gives rise to infinite
avalanches, with non-zero probability. Then, with probability
$1$, such backgrounds will lead to an infinite avalanche for some
finite value of $N$.
In higher dimensions also, a similar  behavior is expected.

In the models with deterministic relaxation rules there is no simple
quantifier like critical density $\rho^\star$, separating the explosive
and non-explosive backgrounds.  Whether the periodic or random
background is explosive or not depends in a complicated way on the
built-in height correlations. For example consider the
BTW model on a square lattice, where the steady state density
$\rho^\star=17/8=2.125$ \cite{sasm}. It has been shown that a background with a random
assignment of height $3$ with probability $\epsilon$, on a sea of
constant height $2$ is explosive, even for arbitrary small value of
$\epsilon$ \cite{explosion}, although the average density
$\rho_{o}=2+\epsilon$ is much less than $\rho^\star$. On the other
hand, it is also possible
to construct a robust periodic background with density arbitrarily
close to maximum stable height $3$ \cite{explosion}.

We will show that  there is  a large class of backgrounds, with a
range of  densities, for which the  growth is less than  explosive,
but more than   compact. 

\begin{figure}
\begin{center}
\includegraphics[width=8cm,angle=0]{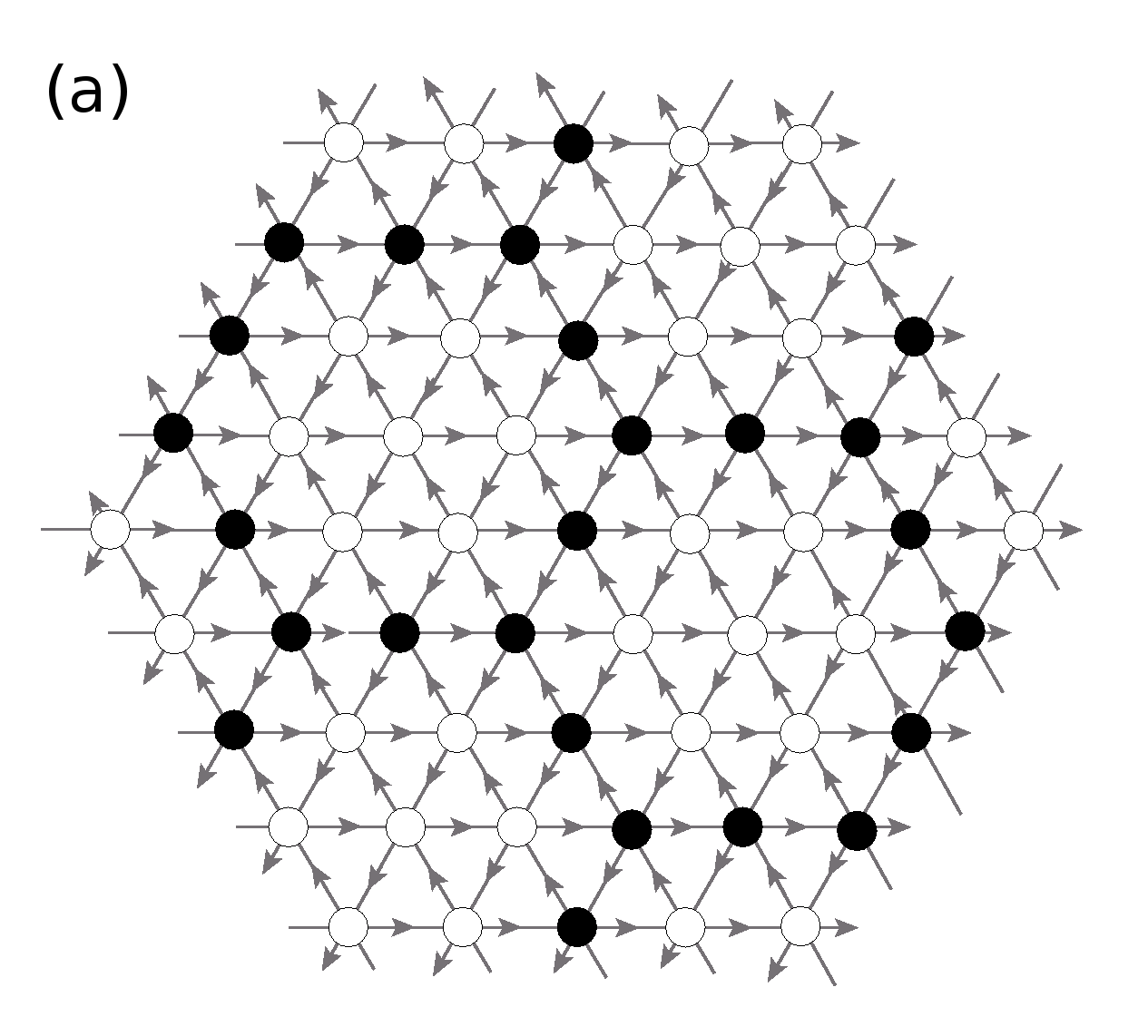}
\includegraphics[width=8cm,angle=0]{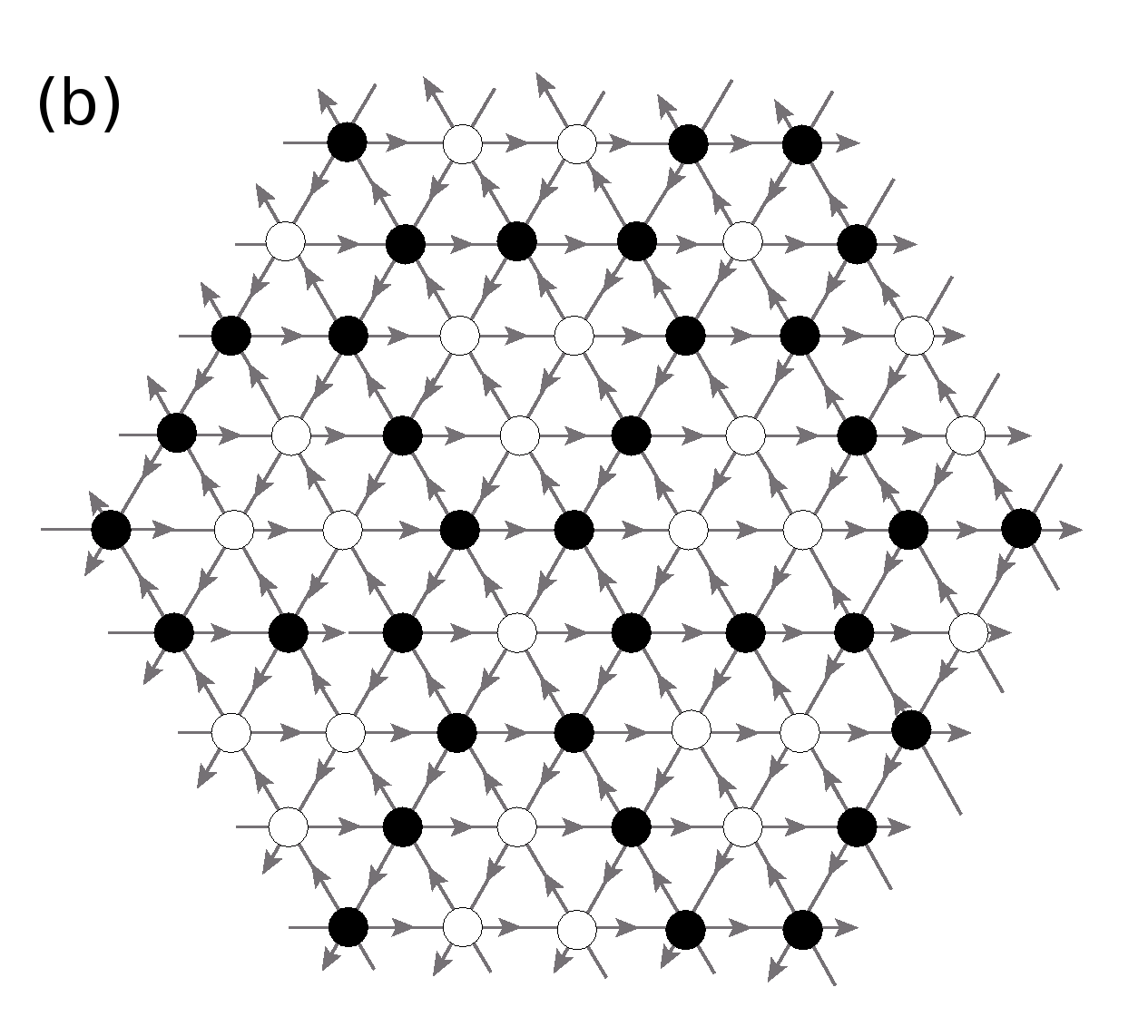}
\caption{Examples of backgrounds of class I and II, respectively. The filled
circles represent height $1$ and unfilled ones height $2$.}
\label{fig:tribg}
\end{center}
\end{figure}
\begin{figure}
\begin{center}
\includegraphics[width=8cm,angle=0]{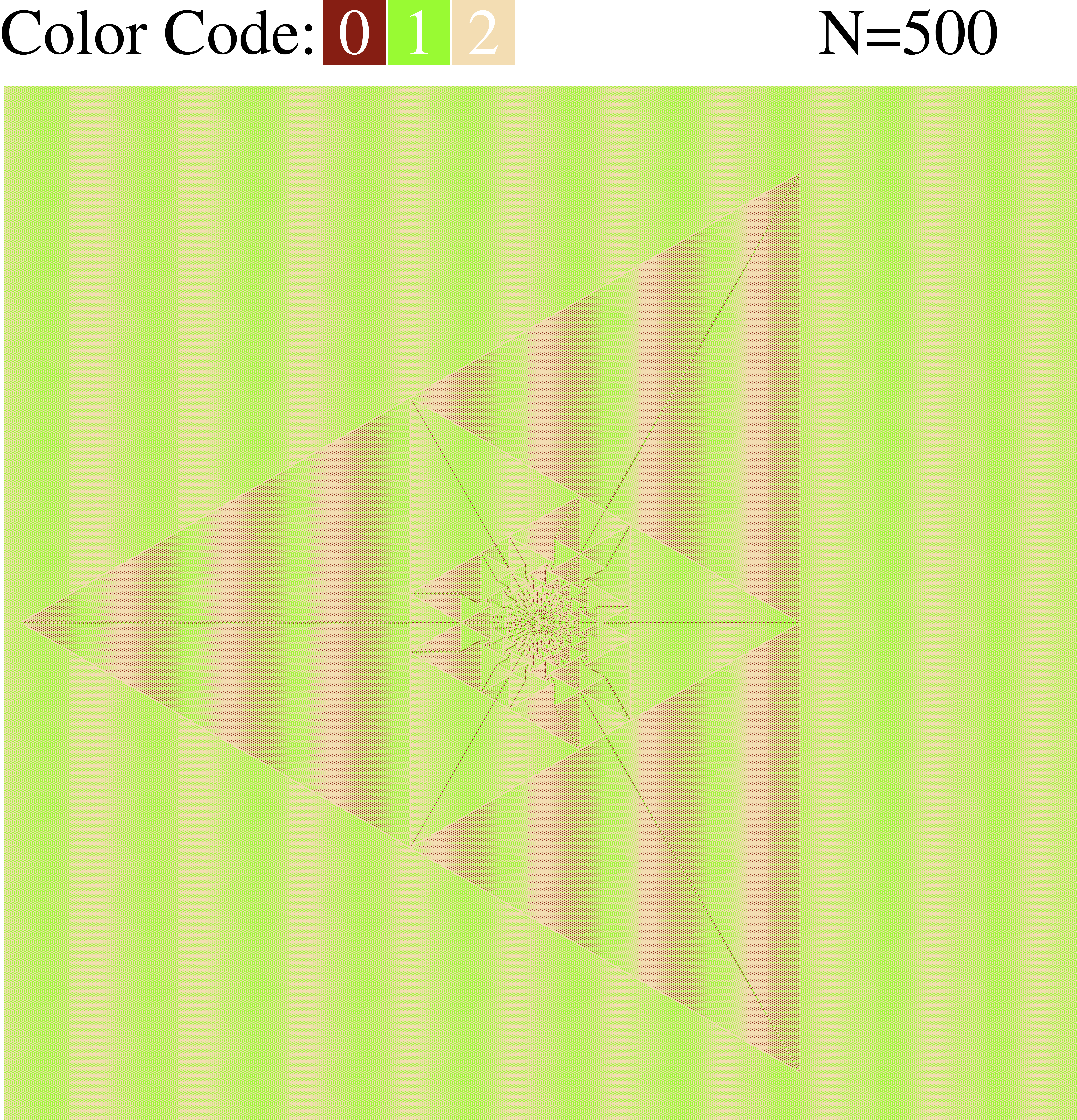}
\caption{(Color online.) The class I pattern, formed by adding $N=500$ particles
at the origin on the first background shown in Fig. \ref{fig:tribg}.
The apparent uniform green color of the background is actually a
periodic structure. Details can be viewed in the electronic
version using zoom in.}
\label{fig:hex}
\end{center}
\end{figure}
\begin{figure}
\begin{center}
\includegraphics[width=8cm,angle=0]{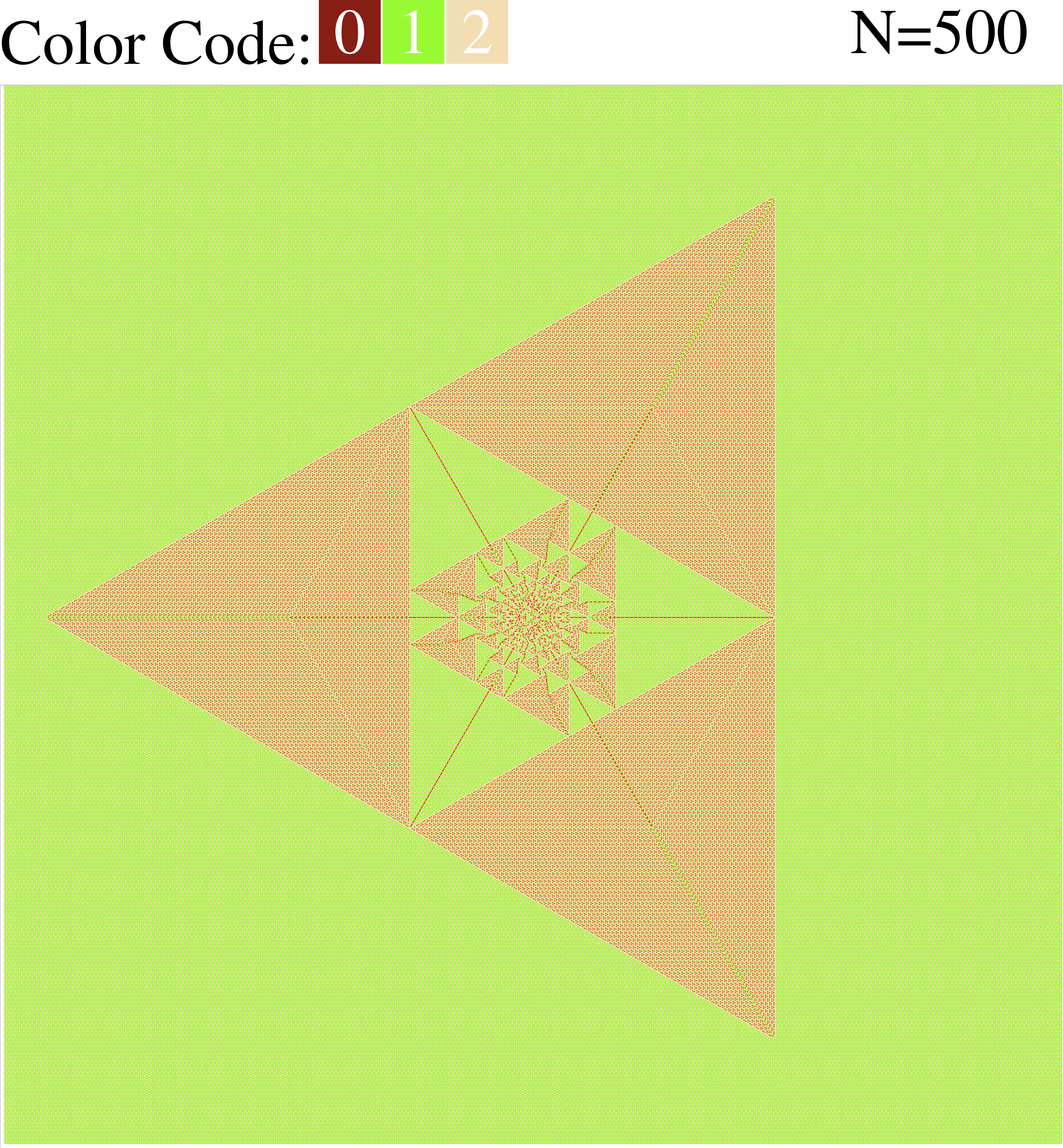}
\caption{(Color online.) The class II pattern, formed  by adding $N=500$ particles
at the origin on the second background in Fig. \ref{fig:tribg}. The
apparent uniform green color of background is actually a periodic
structure. Details can be viewed in the electronic version using zoom
in.}
\label{fig:tri}
\end{center}
\end{figure}

\section{Examples of non-compact growth\label{sec:3}}
We first discuss the patterns with $\alpha =1$. We start with a ASM
on a directed graph corresponding to an infinite two dimensional
triangular lattice, with each site having three incoming and three
outgoing arrows (see Fig. \ref{fig:trilattice}). The threshold
height $z_{c}=3$, for each site. If the height at any site is above or
equal to $z_{c}$, it is unstable, and relaxes by toppling: in each
toppling, three sand grains leave the unstable site, and are
transferred one each along the directed bonds going out of the site.
\begin{figure}
\includegraphics[width=7.8cm]{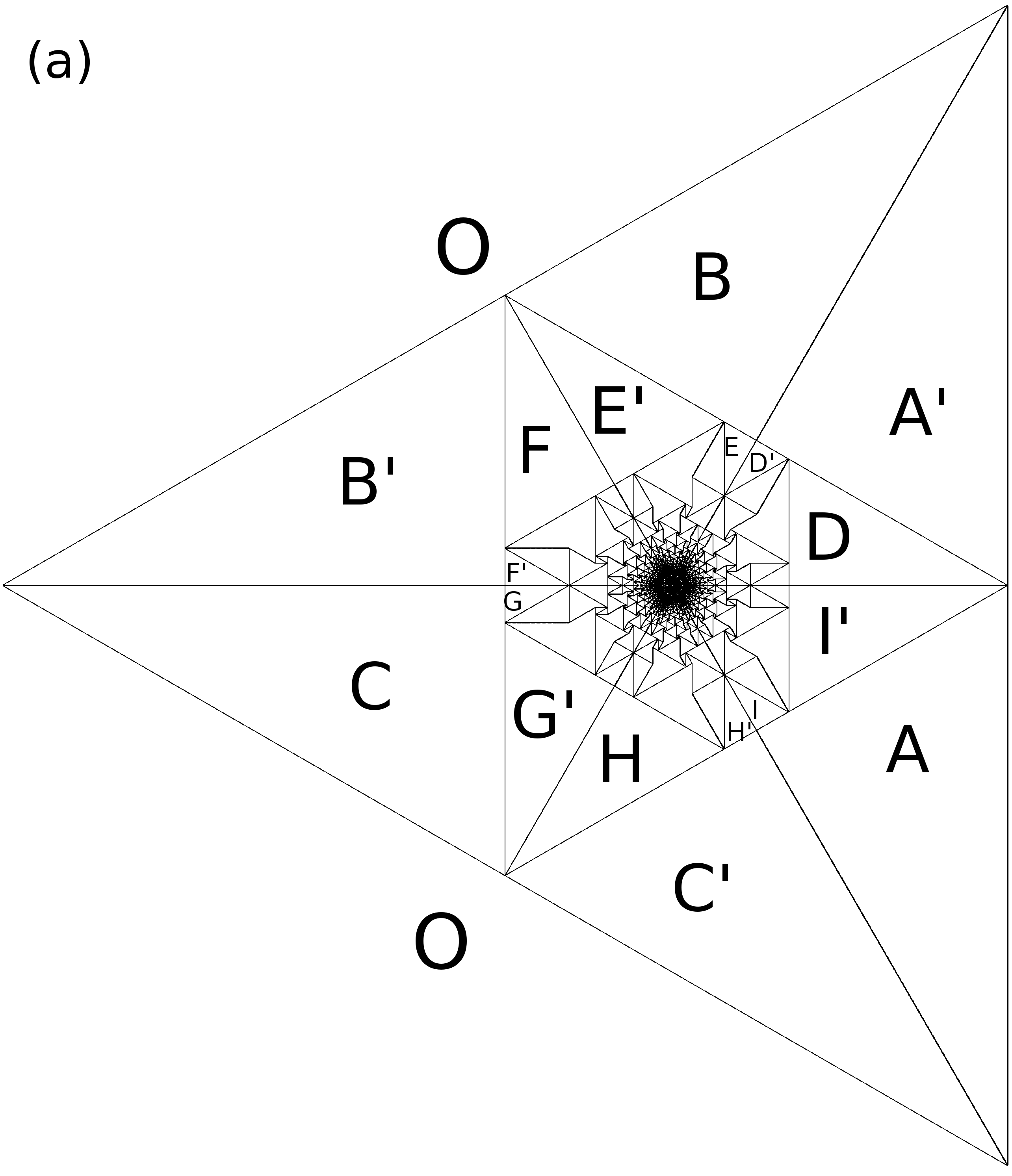}
\includegraphics[width=8cm]{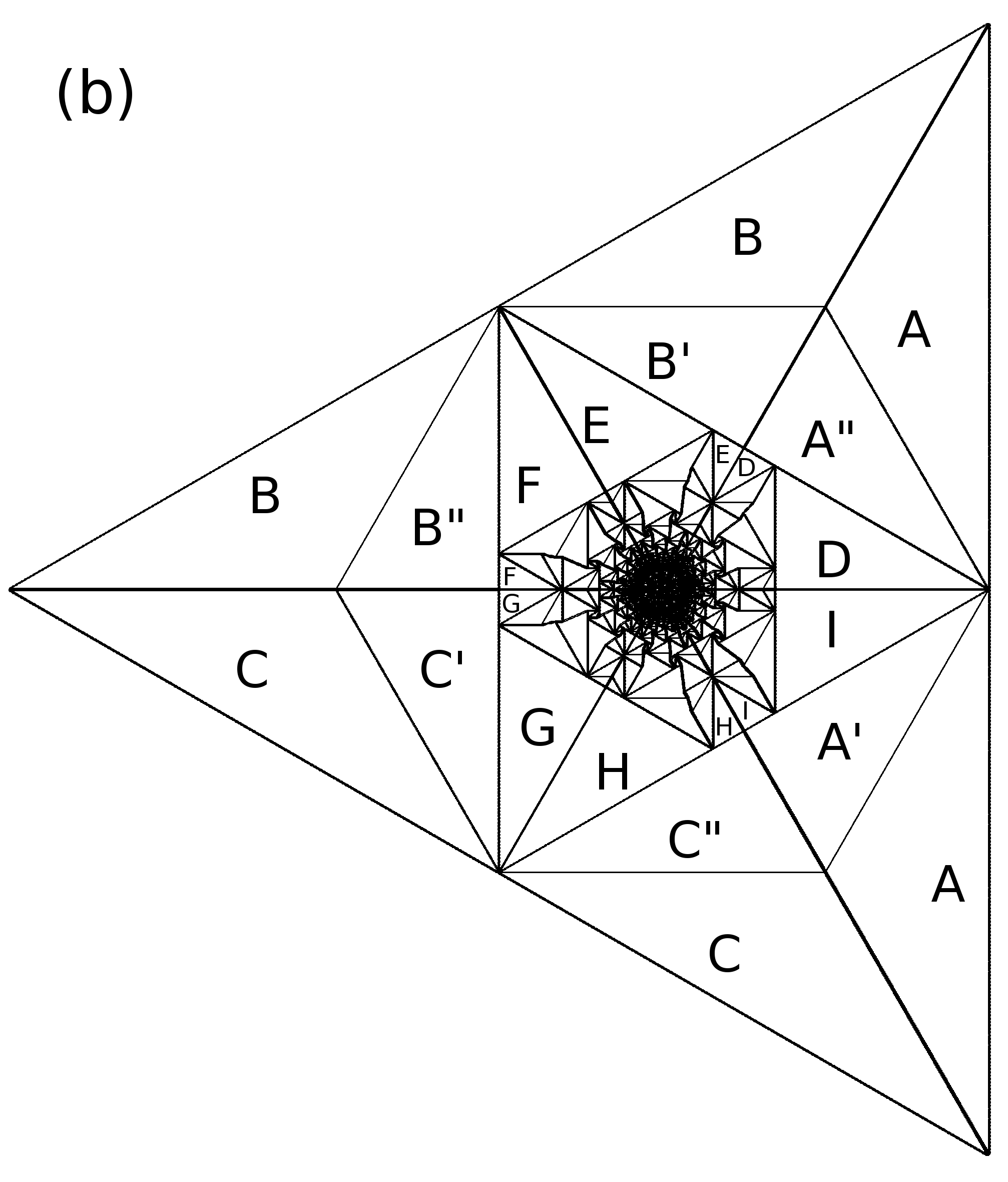}
\caption{The patterns in terms of $Q\left(r\right)$, corresponding
to those in Fig. \ref{fig:hex} and \ref{fig:tri}. Sites with zero
$Q\left(r\right)$ are colored white, and non-zero are colored black.
The larger patches are given identifying labels.}
\label{fig:line}
\end{figure}

We consider two classes of backgrounds on this lattice:
\begin{itemize}
\item[]\underline{\textbf{Class I:}} We consider the lattice as made of triangular plaquettes, which are joined together to make tiles in the shape of regular hexagons  with edges of length $l$. We cover the two-dimensional plane with these tiles.  Sites that lie on the boundaries of these hexagons are assigned height $1$, and the rest of the sites have height $2$.  Figure 
\ref{fig:tribg}$(a)$ shows the background configuration for the case $l=2$. 

\item[]\underline{\textbf{Class II:}}
For these backgrounds, we cover the two-dimensional plane with tiles
in the shape of equilateral triangles of edge-length $l$. The sites
that lie on the boundaries of the triangles, and are shared by two
triangles, are assigned  height  $1$, and remaining sites are assigned
height $2$. Sites that are at the corners of triangular tiles, and
shared by six of them, are also assigned heights $2$. The background
configuration corresponding to $l=4$ is shown in figure \ref{fig:tribg}$(b)$. The pattern
made of triangular tiles with $l=3$ is same as the class I background
with hexagon of edge-length $1$. Hence, only patterns formed  with
triangles of edge-length $ l \ge 4$ will be said to be in this class.
\end{itemize}

The patterns produced by adding $N$ grains, where $N$ is large, at a single
site on the two backgrounds in Fig. \ref{fig:tribg} are shown in
Fig. \ref{fig:hex} and \ref{fig:tri}. While the patterns look quite
similar, a closer examination shows that they are not identical. In
Fig. \ref{fig:tri}, there are extra lines of particles within the
brownish patches which break each patch into smaller parts. In fact,
with the identification of some patches having only a point in common,
as discussed later, we can show that each patch breaks into exactly
three patches. These three parts have similar periodic pattern, but
with different orientations.

This differences can be seen more clearly in terms of the net excess
change in height ${Q\left( R \right)}$ in a unit cell centered at
$R$, where the unit cell is that of the background pattern.
\begin{equation}
Q\left( R \right)=\sum_{R'\in \textrm{unit cell}}\Delta z(R+R'),
\end{equation}
where $\Delta z\left( R \right)$ is the change in height at site $R$.
For example in the first background in Fig. \ref{fig:tribg} a unit
cell is a hexagon of edge length $l=2$, and for the second background
it is a parallelogram of each side length $l=4$. A site that is on the edge of
the unit cell is counted with weight $1/2$, and a site on the corner
of the hexagon with weight $1/3$, and on the corner of the
parallelogram with weight $1/4$. By construction, the function
$Q\left(r\right)$ is zero inside each patch, and non-zero along the
boundaries between patches. The patterns in terms of these variables,
corresponding to those in Fig. \ref{fig:hex} and \ref{fig:tri} are
shown in Fig. \ref{fig:line}.
\begin{figure}
\begin{center}
\includegraphics[scale=0.70,angle=0]{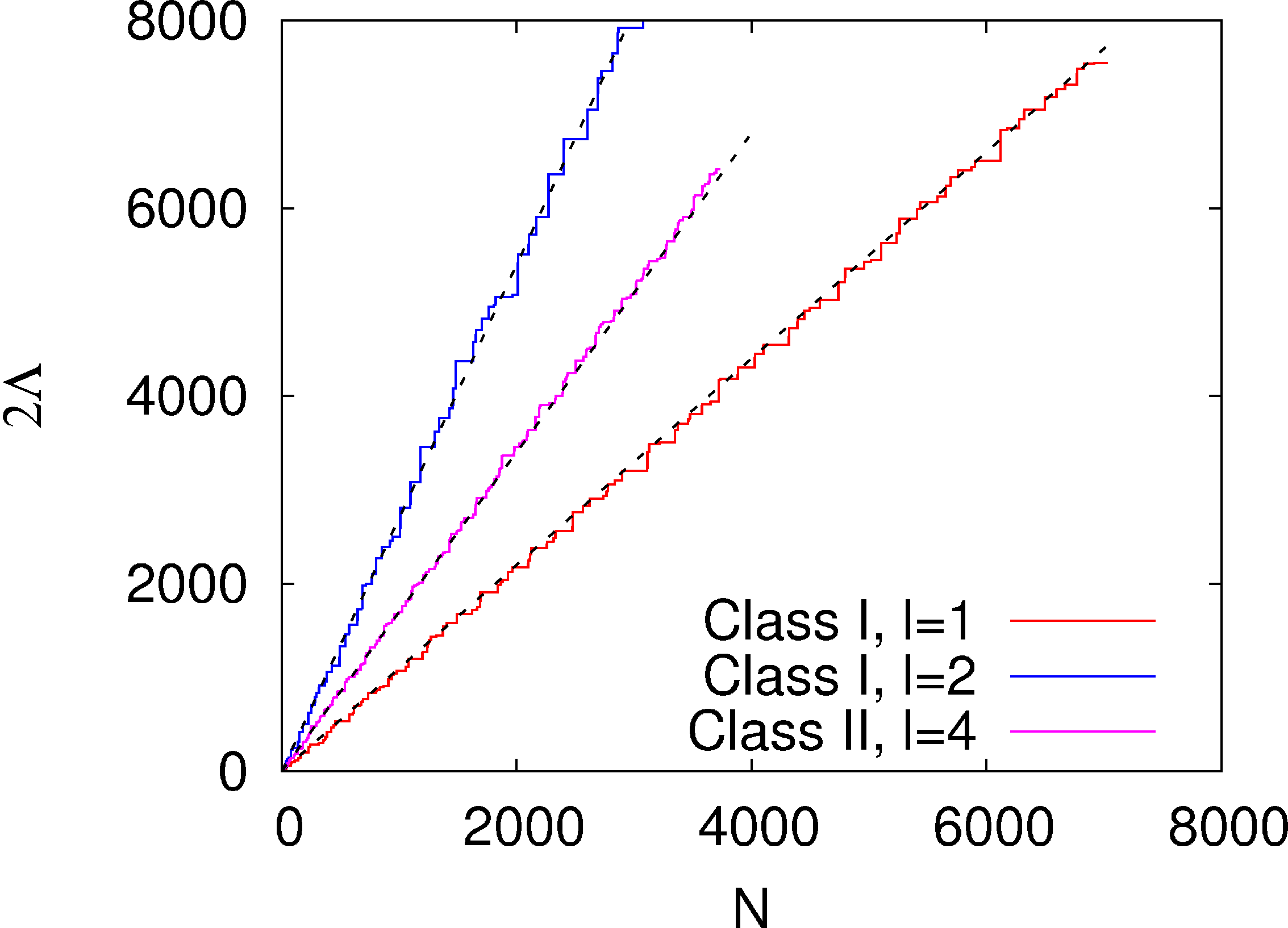}
\caption{(Color online.) The diameter $2\Lambda$ of the patterns as a function of the
number $N$ of  added  grains. The cases shown are  (i) Class I, $l=1$,
(ii) Class I, $l= 2$, and (iii) Class II , $l=4$. The corresponding
straight line fits have slopes given by $1.1$, $2.7$ and $1.7$ respectively.}
\label{fig:triln}
\end{center}
\end{figure}

We have seen that the patterns on these two classes of backgrounds
exhibit proportionate growth, \textit{i.e.}, all the spatial features
inside the patterns for large $N$, grow at the same rate with the
diameter. We define the diameter $2\Lambda$, in general, for any pattern in
this paper, as the height of the smallest rectangle containing it. For
the patterns in Fig. \ref{fig:hex} and Fig. \ref{fig:tri}, it is then
the length of a side of the bounding
equilateral triangle.
This particular choice makes $2\Lambda$ as
an integer multiple of $\sqrt{3}$, on the triangular lattice. We find that for both types of backgrounds
in Fig. \ref{fig:tribg}, the diameter of the pattern grows linearly
with $N$ (Fig. \ref{fig:triln}).

\section{Piece-wise linearity of the toppling function.\label{sec:4}}
Considering the proportionate growth, let us define a rescaled
coordinate $\vec r=\vec R /N^{\alpha}$,
where $\vec R\equiv\left( x,y \right)$ is the position vector of a
site on the lattice.
The number of topplings at any site inside the pattern, scales linearly with
$N$. Let us define
\begin{equation}
\phi\left( \vec r \right)=\lim_{N\rightarrow
\infty}\frac{T_{N}( \vec R )}{N}.
\end{equation}
We now show, using an extension of the argument given in \cite{myepl},
that the function $\phi$ is linear inside periodic patches
in all the patterns with non-compact growth, \textit{i.e.}, with
$\alpha>1/2$. Within a patch, the function
$\phi(\vec r)$ is expandable in Taylor series around any point
$\vec r_{o}$, not on the boundary of the patch. Defining $\vec r_{o}\equiv\left( \xi_{o},
\eta_{o} \right)$, and
$\Delta\vec r_{o}\equiv\left( \Delta\xi, \Delta\eta \right)$ we have
\begin{eqnarray}
\phi\left( \xi_{o}+ \right.&&\Delta \xi, \left.\eta_{o}+\Delta\eta \right)-\phi\left(
\xi_{o}, \eta_{o} \right)\nonumber\\
&&=d\Delta\xi+e\Delta\eta+\mathcal{O}\left(
\Delta\xi^{2},\Delta\eta^{2},\Delta\xi\Delta\eta \right).
\end{eqnarray}
Consider any term of order $\ge 2$ in the expansion, for example, the term
$\sim(\Delta\xi)^2$. This can only arise due to a term $\sim (\Delta
x)^2N^{1-2\alpha}$ in the toppling function $T_{N}(
\vec R)$.
Then, considering the fact that $T_{N}(
\vec R )$ is an integer function of $x$ and $y$, it is easy
to see that this term would lead to discontinuous changes in
$T_{N}(\vec R)$ at intervals of $\Delta x \sim
\mathcal{O}(N^{\alpha-1/2})$. As $\alpha>1/2$ for non-compact growth
patterns, this leads to a change in the periodicity of heights at such intervals
inside each patch which themselves are of size $\sim N^{\alpha}$. This
would then result in many defect lines within a patch, in the pattern at large $N$. However
there are no such features in Fig. \ref{fig:hexpicl1}.
Therefore inside each periodic patch,
$\phi(\vec r)$ must be exactly linear in $\vec r$. In fact, it turns out
that the integer toppling function $T_{N}(\vec R)$ is exactly linear inside
a patch even for any finite $N$, except for an additional periodic term of
periodicity equal to that of the heights inside the patch.

Another consequence of the  exact linearity of the potential function in each
patch is that all patch boundaries in the asymptotic pattern  are straight lines.

The argument finally relies on the two observed (not rigorously
established) features of the
patterns, \textit{i.e.}, there is proportionate growth, and that the
patterns can be decomposed in terms of periodic patches which are
themselves of size $\mathcal{O}\left( N^{\alpha} \right)$.

Let us write the toppling function $T_{N}(\vec R)$ within a single
patch $P$ as
\begin{equation}
T_{N}( \vec R )=A_P+\vec K_{P}\cdot
\vec R+T_{periodic}( \vec R ),
\label{eq:intT}
\end{equation}
where $T_{periodic}( \vec R )$ is a periodic function of
its argument with zero mean value. If $\hat{e}_{1}$ and
$\hat{e}_{2}$ are the basis vectors at the unit cell of the periodic
pattern then we have
\begin{eqnarray}
T_{N}( \vec R+\hat{e}_{1} )-T_{N}( \vec R)&=&\vec K_{P}\cdot
\hat{e}_{1},\nonumber \\
T_{N}( \vec R+\hat{e}_{2} )-T_{N}( \vec R )&=&\vec K_{P}\cdot
\hat{e}_{2}.
\end{eqnarray}

As $T_{N}( \vec R )$ are integer valued functions,
$\vec K_{P}\cdot \hat{e}_{1}$ and
$\vec K_{P}\cdot\hat{e}_{2}$ can only take integer values. If
$\hat{g}_{1}$ and $\hat{g}_{2}$ are the unit vectors in the reciprocal
space of the super lattice of the periodic pattern,
\begin{equation}
\hat{g}_{i}\cdot\hat{e}_{j}=\delta_{ij}\textrm{~ ~ ~ ~ }i,j=1,2,
\end{equation}
then $\vec K_{P}$ must be an integer linear combination of
$\hat{g}_{1}$ and $\hat{g}_{2}$, and can be written as
\begin{equation}
\vec K_{P}=n_1 \hat{g}_{1}+ n_2 \hat{g}_{2},
\label{eq:recp1}
\end{equation}
where $n_1$ and $n_2$ are some integers. For example, in the background
pattern in Fig. \ref{fig:bkg0}, a choice
of the basis vectors and its reciprocal vectors is
\begin{eqnarray}
\hat{e}_{1}\equiv\left( \frac{3}{2}, \frac{\sqrt{3}}{2}
\right);&&\mathrm{ ~ ~ ~ ~ ~   }
\hat{e}_{2}\equiv\left( \frac{3}{2}, -\frac{\sqrt{3}}{2}  \right)
\nonumber\\
\hat{g}_{1}\equiv\frac{2}{3}\left( -\frac{1}{2},
-\frac{\sqrt{3}}{2} \right);&&\mathrm{ ~ ~ ~ ~ ~   }
\hat{g}_{2}\equiv\frac{2}{3}\left( -\frac{1}{2},\frac{\sqrt{3}}{2} \right).
\label{eq:recp}
\end{eqnarray}
The fact that $K_{_P}$ is constant
inside a patch, implies that the patches can be labeled by the pair
of integers $\left( n_1, n_2 \right)$.

An interesting consequence of this
linear dependence of $T_{N}( \vec R )$ is that there are no transient structures within
the patches. On increasing $N$, if the $A_{_P}$ function
increases, all sites in the patch $P$, except possibly those at the patch
boundaries, undergo same number of additional
topplings.

\section{Characterizing the class I asymptotic patterns \label{sec:5}}
\begin{figure}
\includegraphics[width=8cm,angle=0]{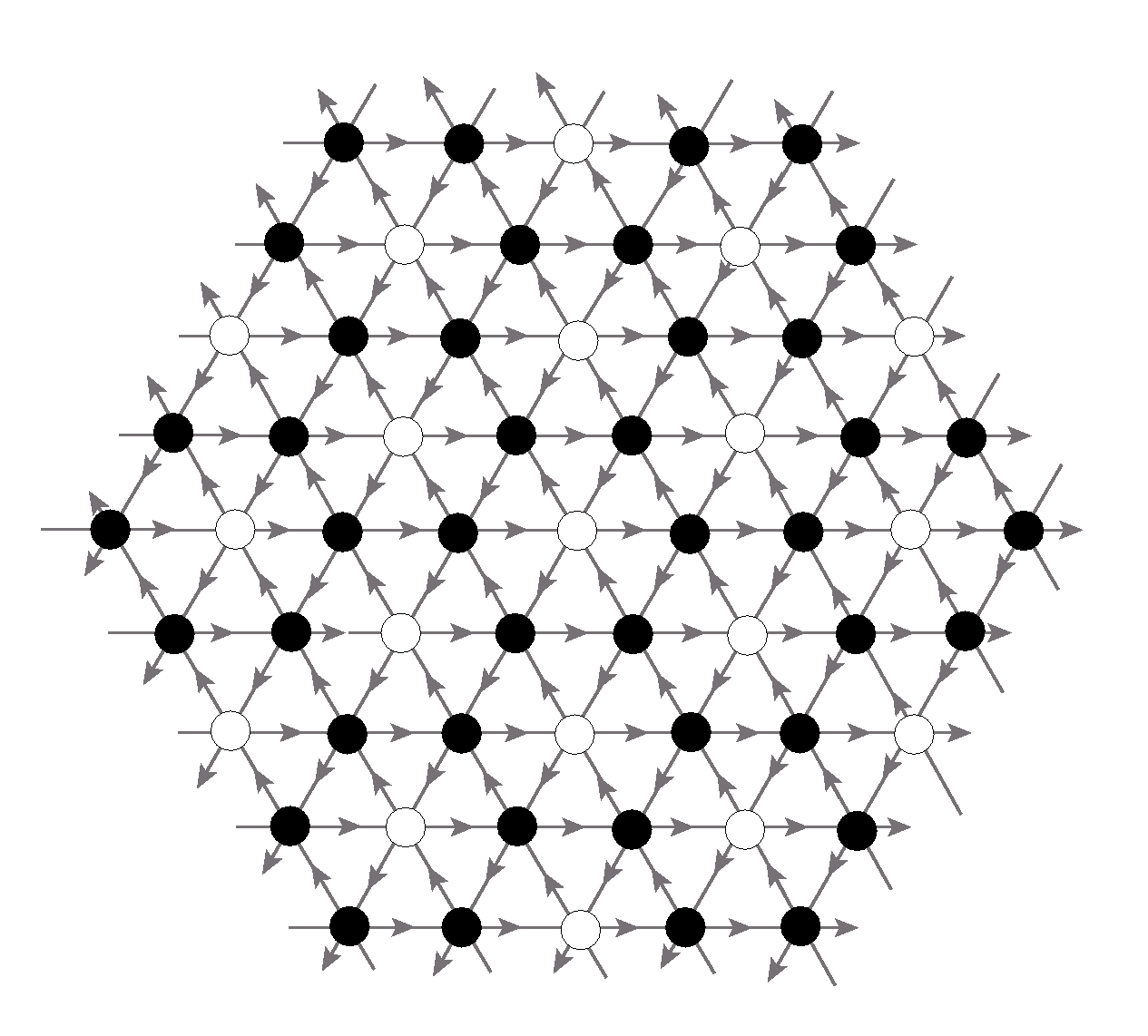}
\caption{The background of class I corresponding to $l=1$. The filled
circles represent height $z=1$ and unfilled ones $z=2$.}
\label{fig:bkg0}
\end{figure}

\begin{figure}
\begin{center}
\includegraphics[width=8cm]{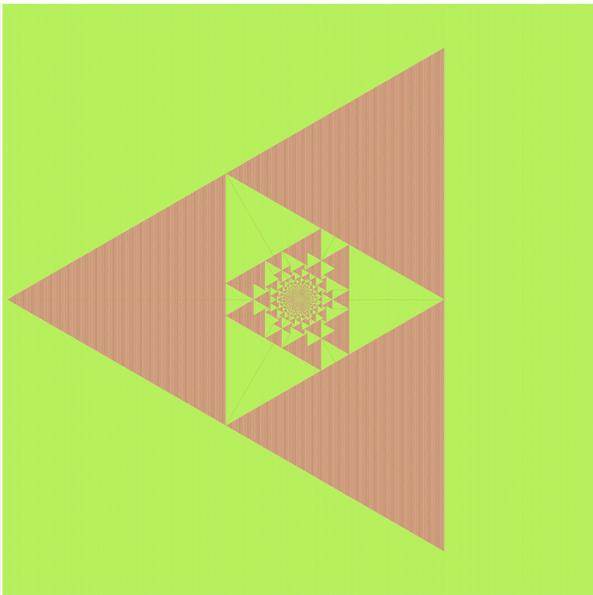}
\caption{(Color online.) The pattern produced by adding $N=3760$ grains at a single site
on the background in Fig. \ref{fig:bkg0}, and relaxing. Details can be
seen in the online version using zoom-in}
\label{fig:hexpicl1}
\end{center}
\end{figure}

We now discuss characterization of the asymptotic pattern of class I,
showing $\alpha =1$. In this section we quantitatively characterize
the asymptotic pattern for the case $l=1$. The background
configuration is shown in Fig. \ref{fig:bkg0}.  A site on the
triangular lattice can be labeled uniquely by a pair of integers
$\left(p, q\right)$, such that its position on a complex plane can be
written as $\mathbf{R}=p+q \omega$, where $\omega=\exp\left( i2\pi/3 \right)$ is
a complex cube root of unity. Then, the height variables in the
background pattern in Fig. \ref{fig:bkg0}, can be written as
\begin{eqnarray}
h(p+q\omega) &=&2 \textrm{ if } p + q =0 \textrm{ (mod 3),}\nonumber\\
  &=&1 \rm{~ ~ ~otherwise.}
\end{eqnarray}
The average height in the background, $\langle z \rangle=4/3$. The
configuration of the pile  produced on this background, by adding
$N=3760$ grains at the origin is shown in Fig. \ref{fig:hexpicl1}.

\begin{figure}
\includegraphics[width=8cm,angle=0]{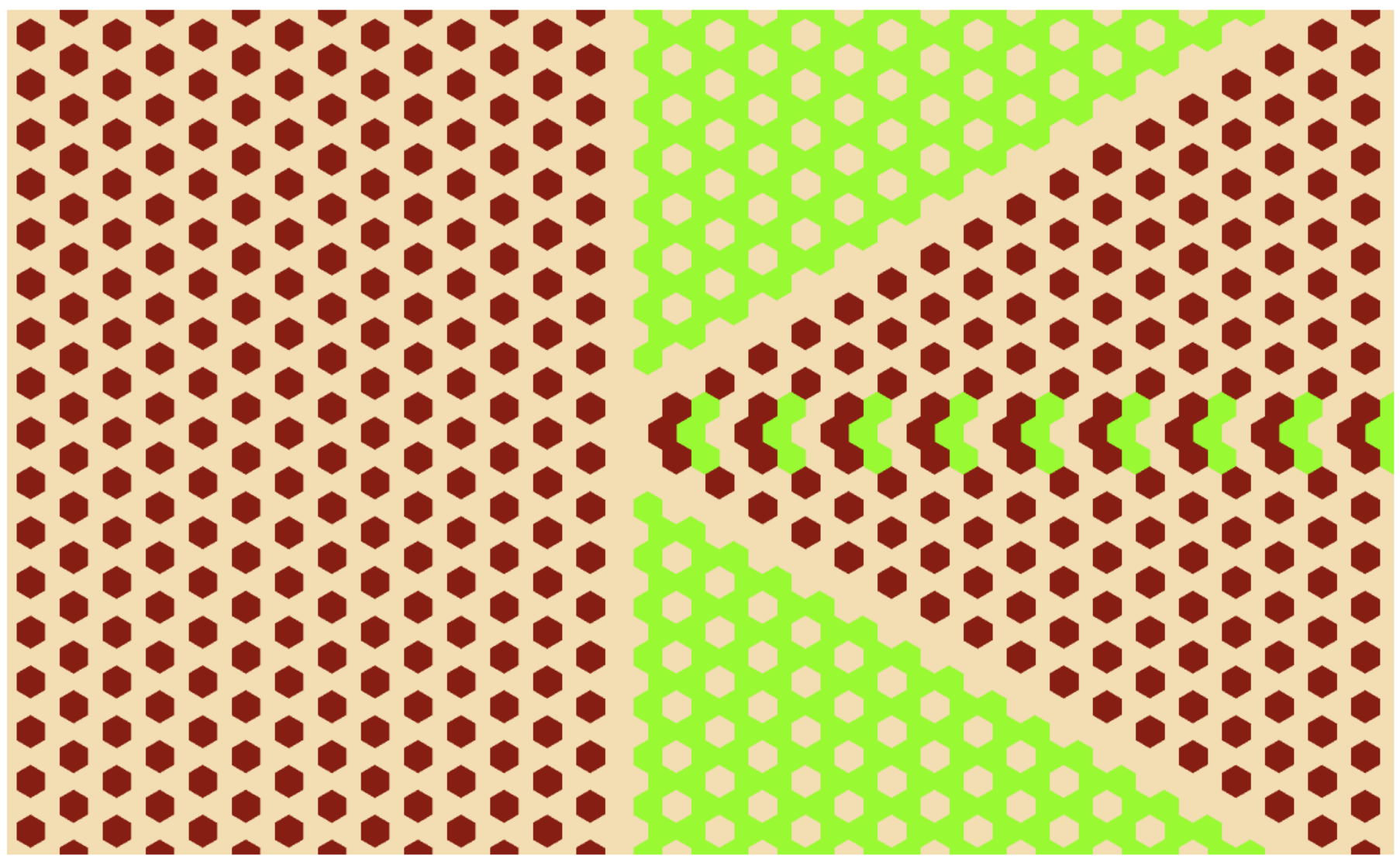}
\caption{(Color online.) An example of patch boundaries in Fig. \ref{fig:hexpicl1}
meeting each other. Each filled hexagon represents Wigner cell around
a site, and the color in them denotes height of that site. The color code is same as in
Fig. \ref{fig:hexpicl1}.}
\label{boundary}
\end{figure}

We see that the sites toppled due to addition of the grains are
confined within an equilateral triangle. The pattern can be thought of
as a union of patches, inside which the heights are periodic. A
zoom-in showing the height configuration with five patches meeting at
a point is shown in Fig. \ref{boundary}. There are only two types of
periodic patches seen: one is like the background, where the sites of
height $2$ are surrounded by sites of height $1$, and the other with
heights $0$ surrounded by heights $2$. Then, the average height
$\langle z \rangle$ inside both types of patches are same. In fact,
it is equal to that of the background, $\langle z \rangle=4/3$.

The patches in the outer region of the pattern are big, and they
become smaller, and more numerous as we go inwards. Along the common
boundary of adjacent patches, we see line-like defect structures, and
only along these lines the density is different from the background. In Fig. \ref{boundary}, one
can also see the periodicity of the structures along the patch
boundaries.  Some patch boundaries, like the horizontal boundary in
Fig. \ref{boundary}, have a deficit of particles compared to the background.

The boundaries of the patches are seen more clearly in terms of
$Q(\mathbf{R})$ variables, as shown in figure
\ref{fig:line}(a), where we have labelled different patches as
$\mathbf{A, A', B, B'}...$ \textit{etc}..

\begin{figure}
\includegraphics[width=8cm,angle=0]{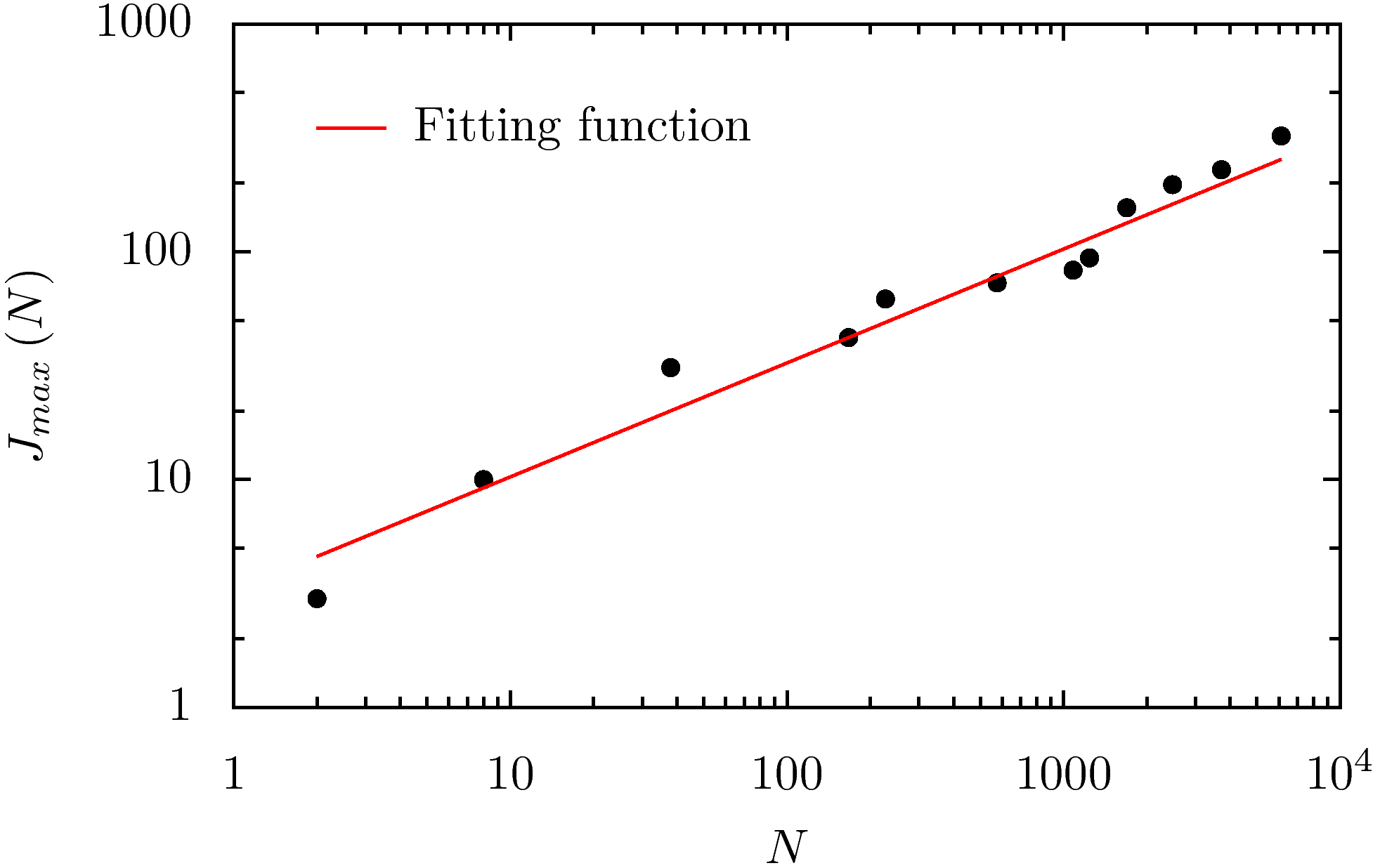}
\caption{$J_{max}(N)$ for the background in Fig.
\ref{fig:bkg0} has a square root dependence on $N$ with a fitting function $3.25\sqrt{x}$.}.
\label{jump}
\end{figure}

The dependence of $2\Lambda$ on $N$ for this background is shown in
Fig. \ref{fig:triln}. We see that the diameter for the pattern  grows
asymptotically linearly with $N$, but it grows in bursts: it remains constant for a
long interval as more and more grains are added, and suddenly
increases by a large amount at certain values of $N$. For example, at
$N=3721$,
the $2\Lambda$ is $2276 \sqrt{3}$, and it jumps to a value
$2408 \sqrt{3}$ when one more
grain is added. Let $J_{max}(N_m)$ denote the size of the
maximum jump in $2\Lambda$ encountered, as $N$ is varied from $1$ to
$N_m$. In Fig. \ref{jump}, we have plotted the variation of $J_{max}(N_m)$ with $N_m$. The graph is consistent with a
power-law  growth, with a power around $1/2$. Thus the fractional size
of the bursts decreases for large $N$.
\begin{figure}
\includegraphics[width=8cm,angle=0]{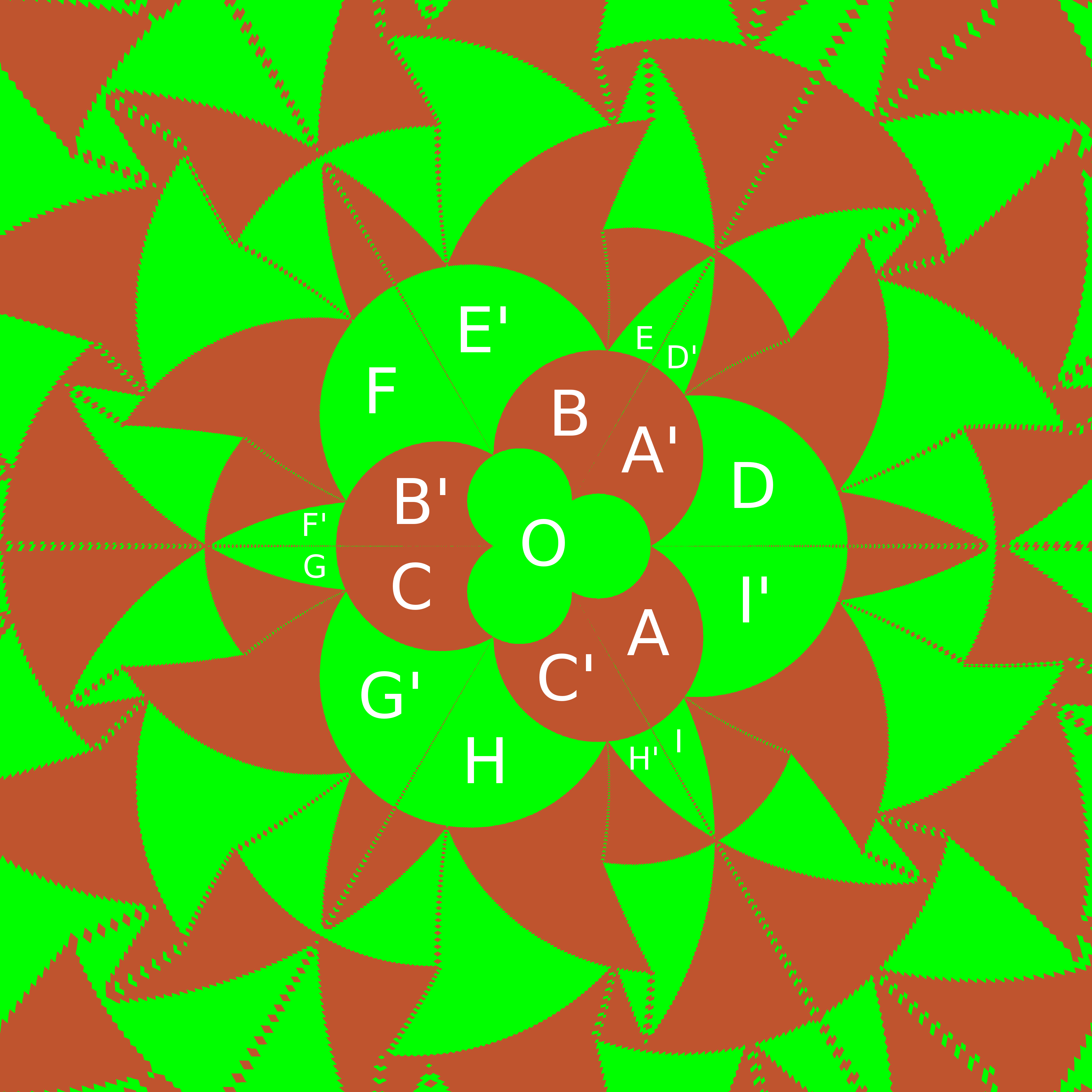}
\caption{The $1/\mathbf{\bar{R}}$ transformation of the pattern in
Fig. \ref{fig:hexpicl1}, where $\mathbf{\overline{R}}$ is the complex
conjugate of $\mathbf{R}$. Labels are the same as used in
Fig. \ref{fig:line}(a).}
\label{fig:1byz}
\end{figure}
\begin{figure}
\includegraphics[width=8cm,angle=0]{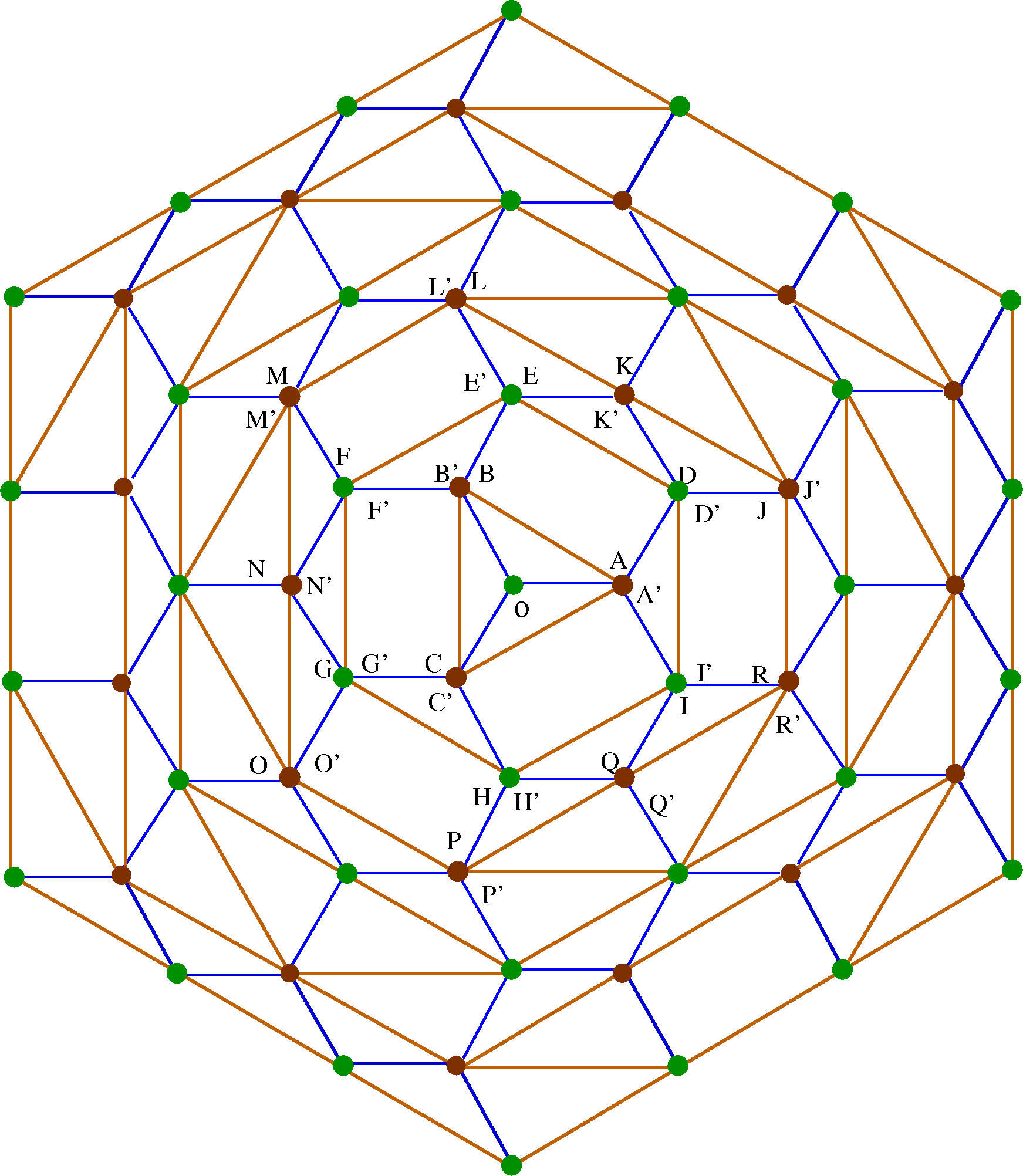}
\caption{The adjacency graph of patches in the pattern in
Fig. \ref{fig:hexpicl1}. The vertices corresponding
to the brownish and greenish patches in the pattern are denoted by
different colors. The pair of patches labeled
by the alphabets and its corresponding primed alphabets in Fig.
\ref{fig:line}(a) are
represented by same vertex on the graph.}
\label{adjhex}
\end{figure}

We define scaled complex coordinates $\mathbf{r}=\mathbf{R}/N$, where
$\mathbf{R}=p+q\omega$ is the complex coordinate of the site
$\left( p,q \right)$. We define the rescaled toppling function for
this pattern as 
\begin{equation}
\phi\left( \mathbf{r} \right)=\lim_{N\rightarrow
\infty}\frac{\sqrt{3}T_{N}\left( \mathbf{r}N \right)}{2 N}.
\label{eq:rscT}
\end{equation}
Then it is easy to see that $\nabla \phi=\left( \partial_{\xi}\phi,
\partial_{\eta}\phi \right)$ is equal to the mean flux of particles at
$\mathbf{r}$. If we consider a small line element
$d\mathbf{l}\equiv\left(d\xi, d\eta\right)$, then the net flux of particles across the line
$d\mathbf{l}$ equals $N \nabla\phi\cdot d\mathbf{l}$. Then, the
conservation of sand grains implies that the toppling function
$T_{N}\left( \mathbf{R} \right)$ satisfies the equation
\begin{equation}
\nabla_{\circ}^{2}T_{N}\left( \mathbf{R} \right)=\delta z\left(
\mathbf{R}\right)-N\delta\left( \mathbf{R} \right),
\end{equation}
where $\nabla_{\circ}^{2}$ is the finite-difference operator on the
lattice, corresponding to the Laplacian $\nabla^{2}$. It is easy to
see that this implies that the scaled potential function $\phi$
satisfies the Poisson equation
\begin{equation}
\nabla^{2}\phi \left( \mathbf{r} \right)=\Delta \rho\left(
\mathbf{r}\right)-\delta\left( \mathbf{r} \right),
\end{equation}
where $\Delta\rho\left( \mathbf{r} \right)$ is the areal density of excess grains at
$\mathbf{r}$. It is related to $\langle\Delta z\left( \mathbf{r}
\right)\rangle$, the mean excess grain density {\it per site} by
\begin{equation}
\Delta \rho\left( \mathbf{r} \right)= \frac{2}{\sqrt{3}} \langle \Delta
z\left( \mathbf{r} \right)\rangle.
\end{equation}

The piece-wise linearity of $\phi$ simplifies the analysis of the pattern,
significantly. The potential function can be characterized by only
three parameters.
Using Eq. (\ref{eq:recp1}), (\ref{eq:recp}) and (\ref{eq:rscT}), for each patch
$P$, we can find a pair of integers $\left( m,n \right)$ such that the
potential in patch $P$ is characterized by
\begin{equation}
\phi\left( \mathbf{r} \right)=- \frac{1}{2\sqrt{3}}\left(
\mathbf{D}_{m,n}\overline{\mathbf{r}}+ \overline{\mathbf{D}}_{m,n}r
\right)+f_{m,n},
\end{equation}
where
\begin{equation}
\mathbf{D}_{m,n}=m+n\omega,
\label{eq:Dunn}
\end{equation}
and $f_{m,n}$ is a real number, constant everywhere inside the patch.
Here $\overline{z}$ denotes the complex conjugate of $z$.

Each patch is characterized by a complex number $\mathbf{D}_{m,n}$ which is the
coefficient  in the potential function $\phi\left( \mathbf{r} \right)$ of the patch. 
In the complex $\mathbf{D}$-plane, each patch with labels as in Fig. \ref{fig:line}(a) can then  be represented by a point. We
connect two patches by a line if they share a common boundary.
Then the resulting figure, shown in Fig. \ref{adjhex}, is the
adjacency graph of the patches. 

We can determine the connectivity structure of this graph, without
knowing the full potential function in each patch. We first take
$1/\mathbf{\bar{R}}$
transformation of the pattern. This is shown in Fig. \ref{fig:1byz}. 
Some of the bigger patches are denoted by capital alphabets in
Fig. \ref{fig:line}(a) and their corresponding patches on the
transformed pattern in Fig. \ref{fig:1byz}.
The patches $\textbf{A}$ and $\textbf{A}'$ in Fig. \ref{fig:line}(a) are adjacent
to the outer region $\textbf{O}$ through the same vertical boundary.
Matching the values of the function $\phi( \mathbf{r})$ and fixing the discontinuity in its normal derivatives at the boundary,
it is easy to see that $\phi(\mathbf{r})$ has the same functional form in the
patches $\textbf{A}$ and $\textbf{A}'$. In fact, it is convenient to
imagine that the boundary between $\textbf{O}$ and $\textbf{A}$ moved
to the right by an infinitesimal amount, so that it
does not touch the patches $\textbf{D}$ and $\textbf{I}'$, and then
$\textbf{A}$ and $\textbf{A}'$ would actually
join to form a single connected patch $\textbf{A}$. We thus consider
$\textbf{A}$ and $\textbf{A}'$ as one patch, and both can be
represented as one point
on the $\mathbf{D}$-plane. Similarly, we identify $\textbf{B}$ and $\textbf{B}'$, $\textbf{C}$
and $\textbf{C}'$, \textit{etc}. 
Then the adjacency graph can be
constructed by joining the sites on the $\mathbf{D}$-plane, according
to the adjacency of patches in Fig. \ref{fig:1byz}.

It turns out that the patches corresponding to $m+n=2\left(
\right.$mod$\left. 3 \right)$ do not appear in the pattern, and
the adjacency graph, as shown in Fig. \ref{adjhex}, is a hexagonal lattice with some extra edges shown in
brown color. These extra edges connect all the vertices at same distance from the origin $\left( 0,0
\right)$ (in the $L^{1}$ metric), and also connect some of the
diagonally opposite sites on the rectangular faces of the graph as shown in figure.

The charge density $\Delta\rho\left( \mathbf{r} \right)$ is zero
inside the patches, and the excess grains due to addition are
distributed along the patch
boundaries, leading to nonzero line charge densities separating
neighboring patches. Then the density function $\Delta\rho\left(
\mathbf{r}
\right)$ is
a superposition of the line charge densities along the patch
boundaries. There are three kinds of line charges of charge density
$\lambda=-1/\sqrt{3}$, $1$, and $2/\sqrt{3}$.

From the electrostatic analogy, it is seen that $\phi\left(
\mathbf{r} \right)$ is
continuous across the common boundary between neighboring patches, and
its normal derivative is discontinuous by an amount equal to the line
charge density $\lambda$ along the boundary. Let $P$ and $P'$ be the two
neighboring patches with the equation of the boundary between them
\begin{equation}
\mathbf{r}=|\mathbf{r}| \exp\left( i \theta \right)+\mathbf{A},
\end{equation}
such that the patch $P'$ is on the left of the boundary.
Then using the continuity condition, it is easy to show that
\begin{eqnarray}
\mathbf{D}_{p'}-\mathbf{D}_{p}&=& i\lambda\sqrt{3} \exp\left( i\theta \right) \rm{~and~} \nonumber \\
f_{p'}-f_{p}&=&Re[\overline{\mathbf{A}}\left( \mathbf{D}_{p'}-\mathbf{D}_{p} \right)]/\sqrt{3},
\label{eq:bc}
\end{eqnarray}
where $\overline{\mathbf{A}}$ is the complex conjugate of $\mathbf{A}$.
We note that, there are only six different types of patch boundaries in the pattern, with angle $\theta$
an integer multiple of $\pi/6$.

It is easy to check that the matching conditions along the edges of hexagonal lattice
(denoted by blue solid line in Fig. \ref{adjhex}) are sufficient to
determine $D_{m,n}$ for
all the vertices. The line charge density $\lambda=-1/\sqrt{3}$ for
the patch boundaries corresponding
to these edges. 
Also, the potential function $\phi=0$, for the vertex at the origin, and
hence, $D$ and $f$ both vanishes.
Then using the matching condition, it is easy to check that, the values
of $D_{m,n}$ are consistent with the form in Eq. (\ref{eq:Dunn}).

The function $f_{m,n}$ satisfies the discrete Laplace's equation on the underlying hexagonal lattice
of the adjacency graph \textit{i.e.}
\begin{equation}
\sum_{m',n'}f_{m',n'}-3f_{m,n}=0 \textrm{ ~ ~ ~  for }\left( m,n \right)\ne 0,
\label{laplace}
\end{equation}
where $\left( m', n' \right)$ denotes the three neighbors of the vertex $\left(m,n\right)$ on the hexagonal lattice.
This can be checked from the concurrency condition of patch boundaries. For
example consider the edges $\mathbf{OA}$, $\mathbf{DA'}$ and
$\mathbf{I'A}$ on the adjacency graph. The corresponding patch boundaries in the
pattern intersect at the same point (Fig. \ref{fig:line}(a)).
Then it is easy to check using the matching condition in
Eq.(\ref{eq:bc}) that,
\begin{equation}
f_{O}+f_{D}+f_{I}=3f_{A}.
\end{equation}
Similar equations hold for the other vertices.

In the region outside the pattern, where none of the sites toppled, the potential function $\phi\left( z \right)=0$. This corresponds to $m =n =0$, and $f_{0,0}=0$. The solution of the Laplace's equation with the above
boundary condition can be written in the following integral form \cite{atkinson}
\begin{widetext}
\begin{equation}
f_{m,n}=\frac{I}{4\pi^{2}}\int_{-\pi}^{\pi}\int_{-\pi}^{\pi}\frac{1-\cos\left(
k_{1}(2m-n)/3+k_{2}n \right)}{1-\left(
\cos{2k_{2}}+2\cos{k_{1}}\cos{k_{2}} \right)/3}dk_{1}dk_{2},
\rm{~ ~ ~ ~ ~ ~ ~}\textrm{       for $m+n=0$ (mod $ 3$)},
\label{solution}
\end{equation}
\end{widetext}
where $I$ is a normalizing constant, which determines the pattern up to
a scale factor. For the
sites with $m+n=1$ (mod $3$), $f_{m,n}$
are the average of those corresponding to the neighboring sites. As an
example the potential function in region $\textbf{A}$, and $\textbf{C}'$ is
\begin{eqnarray}
\phi_{_{\textbf{A}}}(\mathbf{r})&=&\frac{I}{3}-\frac{\xi}{\sqrt{3}}, \\
\phi_{_{\textbf{C}'}}(\mathbf{r})&=&\frac{I}{3}+\frac{1}{\sqrt{3}}\left( \frac{1}{2}\xi +
\frac{\sqrt{3}}{2}\eta \right),
\end{eqnarray}
where $\mathbf{r}=\xi+i \eta$, and $i=\sqrt{-1}$.
Then the equation of the patch boundary between patches $\textbf{A}$ and $\textbf{O}$ is
\begin{equation}
\xi=I/\sqrt{3},
\label{eq:aob}
\end{equation}
and that of the boundary between patches $\textbf{C}'$ and $\textbf{O}$ is
\begin{equation}
\sqrt{3}\xi+3\eta+2I=0.
\end{equation}
Equivalently, the length of an edge of the bounding
equilateral triangle of the pattern is equal to $2I N$, for large
$N$.

The constant $I$ in Eq. (\ref{solution}) can be calculated using the form of the potential
function near the site of addition.
As noted, the function $\phi$ can be considered as the potential due to
line charges along the patch boundaries and a point
charge of unit amount at the origin. Then, close to the origin the
solution diverges logarithmically as $\widetilde{\phi}\left(
\mathbf{r}
\right)=-\left(2\pi\right)^{-1}\log\left( |\mathbf{r}| \right)$,
and the potential function is an approximation to this solution by a piece-wise linear function.
Then, there are coordinates $\mathbf{r_{o}}$ inside each patch $\left( m,n
\right)$ with $|m|+|n|$ large, where the $\phi$ and its first
derivatives are equal to $\tilde\phi$ and its first derivatives,
respectively. Then,
\begin{eqnarray}
2\sqrt{3}\frac{\partial}{\partial
\overline{\mathbf{r}}}\widetilde{\phi}\left( \mathbf{r}
\right)\vert_{\mathbf{r_{o}}}&\simeq&-\mathbf{D}_{m,n} \rm{~ ~ and ~ ~} \nonumber \\
-\frac{1}{2\sqrt{3}}\left\{
\mathbf{D}_{m,n}\overline{\mathbf{r_{o}}}+\overline{\mathbf{D}}_{m,n}\mathbf{r_{o}}
\right\}+f_{m,n}&\simeq&-\frac{1}{2\pi}\log\left( |\mathbf{r_{o}}|
\right).\nonumber\\
\end{eqnarray}
The above two equations imply
\begin{equation}
f_{m,n}\simeq\frac{1}{2\pi}\log\left( |m+n\omega| \right),
\label{logasymp}
\end{equation}
for $|m|+|n|$ large.
Comparing it with the Eq. (\ref{solution}) for large $|m|+|n|$ we find
that the numerical constant $I=1/\sqrt{3}$. This determines the potential function completely, and thus
characterizes the pattern. For example, as in figure
\ref{fig:line}(a), the equation of the rightmost boundary of the
pattern, using Eq. (\ref{eq:aob}) is $x=N/3$. Equations of other boundaries
of patches can be calculated similarly. For example, the reduced
coordinates of the point where the patches $D$ and $D'$ meet in Fig.
\ref{fig:line}(a), is determined by the condition that it is a
common point of patches $D$, $J$ and $A'$, and that the function
$\phi$ is continuous.
\begin{eqnarray}
f_{1,0}-\frac{1}{\sqrt{3}}\xi &=& f_{2,1}-\frac{1}{\sqrt{3}}\left( \frac{3}{2}\xi +
\frac{\sqrt{3}}{2}\eta \right) \nonumber \\
&=& f_{3,1}-\frac{1}{\sqrt{3}}\left( \frac{5}{2}\xi +
\frac{\sqrt{3}}{2}\eta \right).
\end{eqnarray}
Then using the values $f_{1,0}=1/3\sqrt{3}$, $f_{2,1}=
1/2\sqrt{3}$, and
$f_{3,1}=7/6\sqrt{3}-1/\pi$ \cite{atkinson} we get the reduced coordinates of this point as 
\begin{equation}
\left(\xi,\eta\right) =
\left(\frac{2}{3}-\frac{\sqrt{3}}{\pi},-\frac{1}{3\sqrt{3}}+\frac{1}{\pi}\right).
\end{equation}

\begin{figure}
\begin{center}
\includegraphics[width=8.0cm]{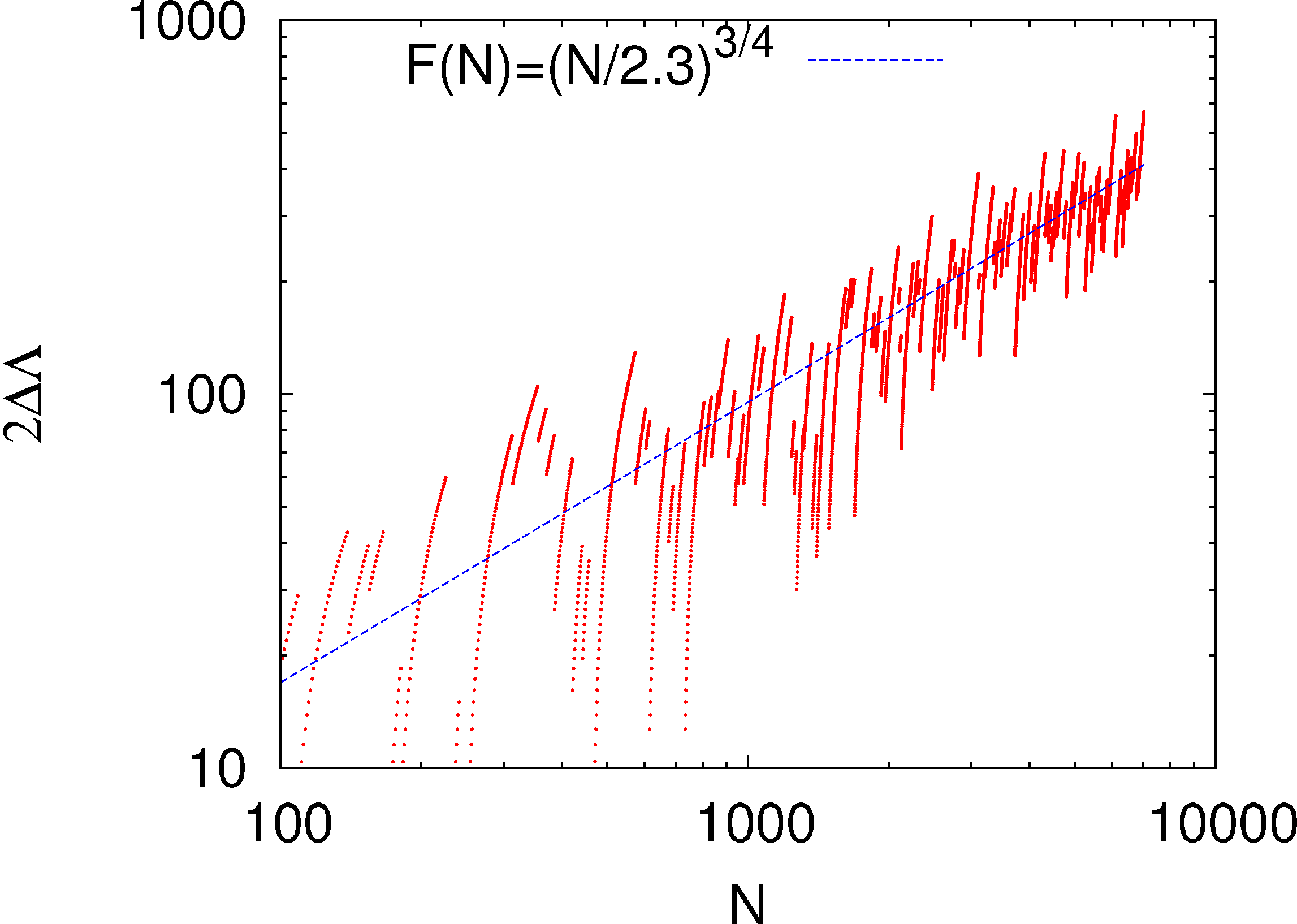}
\caption{The discrepancy $2 \Delta \Lambda$ between the actual height
of bounding triangle, and the asymptotic value $2 N/\sqrt{3}$ plotted
as a function of $N$. The straight line shows a simple power-law fit
with power $3/4$.}
\label{fig:diffdia}
\end{center}
\end{figure}

Equivalently, the height of the bounding equilateral triangle
increases as $ 2 N/\sqrt{3} \simeq 1.154 N$. The estimated slope of
the fitting line in Fig. \ref{fig:triln} is 1.1, in reasonable agreement with the
theory. However,  even though the exact function $\Lambda(N)$  has
large  fluctuations of number theoretic origin, the estimated slope is
noticeably lower than the calculated asymptotic value.  To examine
this discrepancy closer, we have plotted in Fig. \ref{fig:diffdia} the
discrepancy $2 \Delta \Lambda  = 2 N/\sqrt{3} - 2 \Lambda(N)$ as a
function of $N$. We find that this appears to increase with $N$ as
$N^{3/4}$, for large $N$. The reason for this behavior is not
understood yet.

For backgrounds, with $l > 1$, our numerical results suggest that there is a
crossover length $R^\star(l)$, and initially, for $R < R^\star(l)$, the
avalanches grow ``explosively" in size. As a result, the number of
particles  inside a disc of radius $R^\star$ in the final  pattern  is less
than that in the initial background. The net flux of particles going
out of the disc increases with $R$ until the radius becomes of order
$R^\star$. After this, the large-scale properties of the pattern
are the same as that of $l=1$ pattern, with the number of particles added
$A_{\ell} N$, where $A_{\ell}$ is an $\ell$-dependent constant. In
particular, the size of the pattern is $A_{\ell}$ -times the size
of the pattern for $l=1$ with same $N$.  The crossover length
$R^\star$ is expected to grows as $\sqrt{N}$.
\begin{figure}
\includegraphics[width=8.0cm]{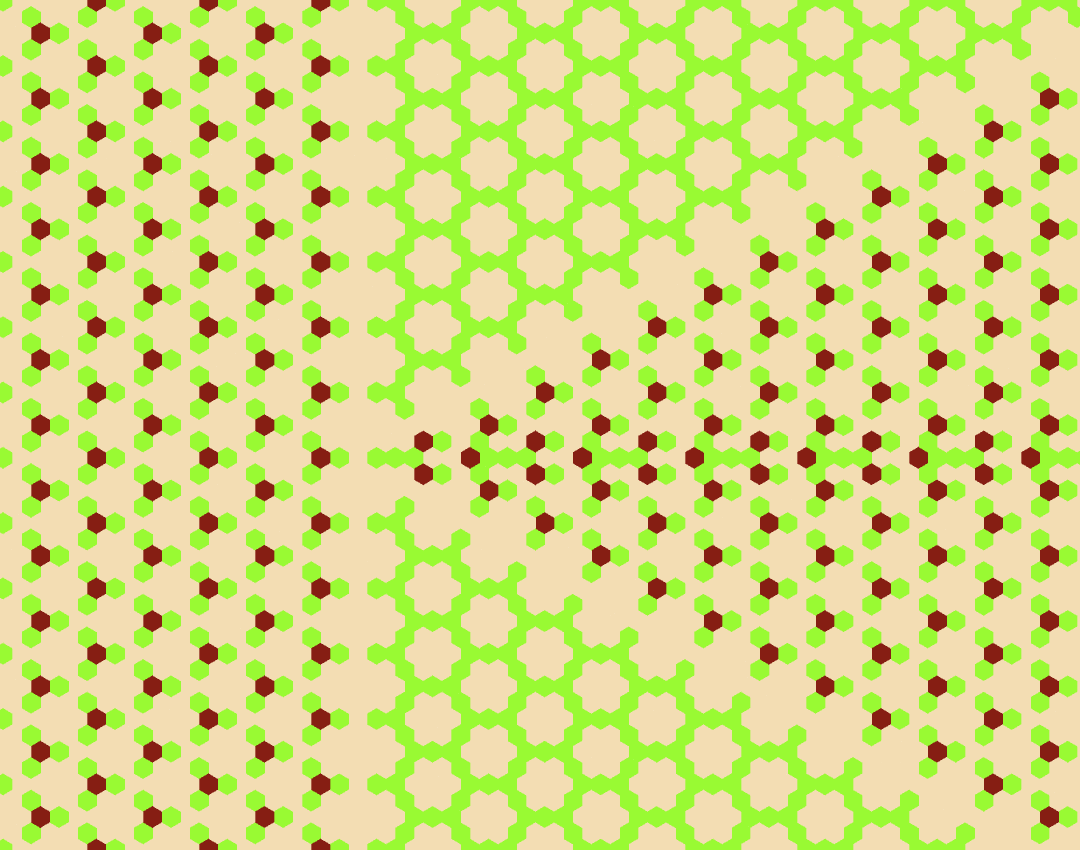}
\caption{An example of five patches meeting at a point, for a
pattern on the background of Class $I$, $\ell=2$. It is easy to check
that the line charge density for the vertical boundary
$\lambda=-1/\sqrt{3}$, same as in Fig. \ref{boundary}. The color code
is same as in Fig. \ref{fig:hexpicl1}.}
\label{fig:pb2}
\end{figure}

For a background with   $\ell > 1$,  the
basis vectors at the unit cell are $\ell\hat{e}_1$ and
$\ell\hat{e}_2$, where $\hat{e}_1$, $\hat{e}_2$ are the basis vectors
for $\ell=1$ background (see Eq. (\ref{eq:recp})). Then the reciprocal basis vectors are
$\hat{g}_1/\ell$ and
$\hat{g}_2/\ell$.  From the observed patterns, we find  that the line charge
densities remain same for any $\ell$ (see Fig. \ref{fig:pb2} for an
example of the patch boundaries).  This implies that $n_1$ and $n_2$ in eq. (8) are constrained to be
multiples of $l$. Writing $n_1 = l m$, $n_2 = l n$, we see that 
the patches can be labeled by the same pair of integers $\left( m,n
\right)$ as in the $\ell=1$ case, and the potential function
$\phi_{(l)}\left( \mathbf{r} \right)$ for general $l$ is related to the $l=1$ case by simple scaling:
\begin{equation}
\phi_{(\ell)}\left( \mathbf{r} \right) =  A_{\ell} \phi_{(1)}\left(
\frac{\mathbf{r} }{A_{\ell} }\right),
\end{equation}
where $A_{\ell}$ is a scale factor. For $\ell=2$, $3$, $4$ and
$5$ the values of $A_\ell$ are approximately $2.34$, $3.38$, $4.41$ and $5.37$,
respectively. We note that $A_\ell$ increases approximately linearly
with $\ell$.

\section{Non-compact patterns with exponent $\alpha < 1$\label{sec:6}}
\begin{figure}
\includegraphics[width=8cm,angle=0]{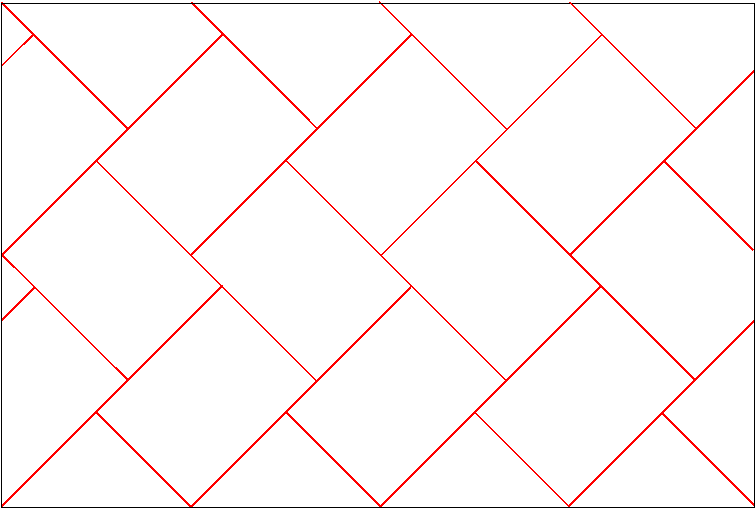}
\caption{A schematic representation of the periodic tiling of the  plane using tilted rectangles. Background 
height patterns with such periodicities on the F-lattice give rise to non-compact growth with the growth-exponent between  $1/2$  and  $ 1$  }
\label{fig:tile_f}
\end{figure}

\begin{figure}
\begin{center}
\includegraphics[width=7cm,angle=0]{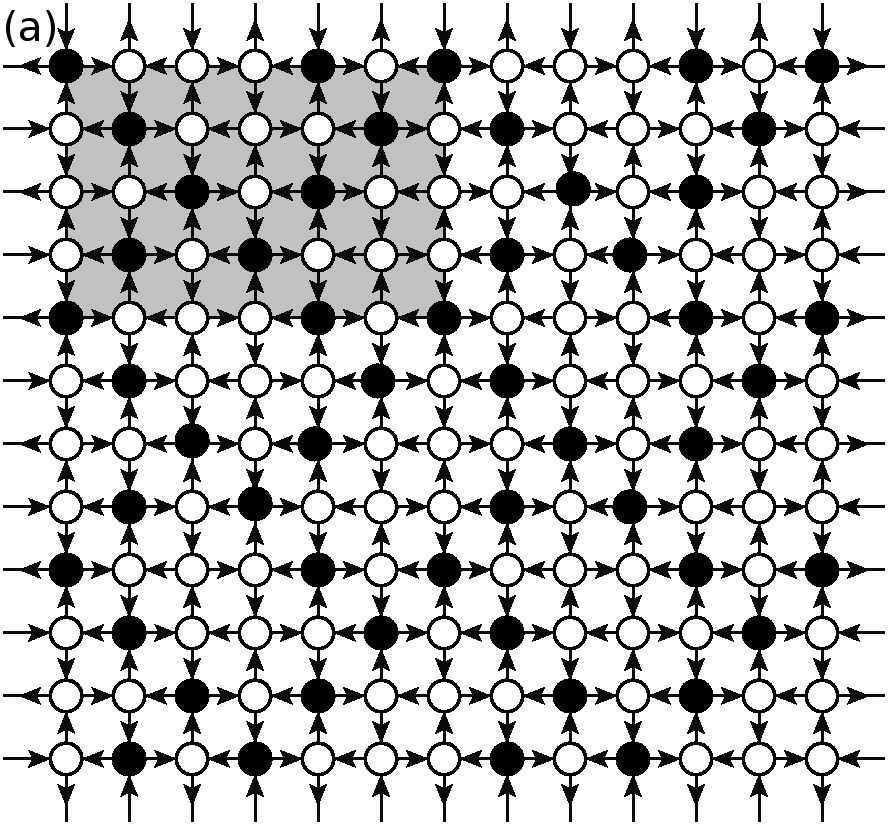}\\
~\\
\includegraphics[width=7cm,angle=0]{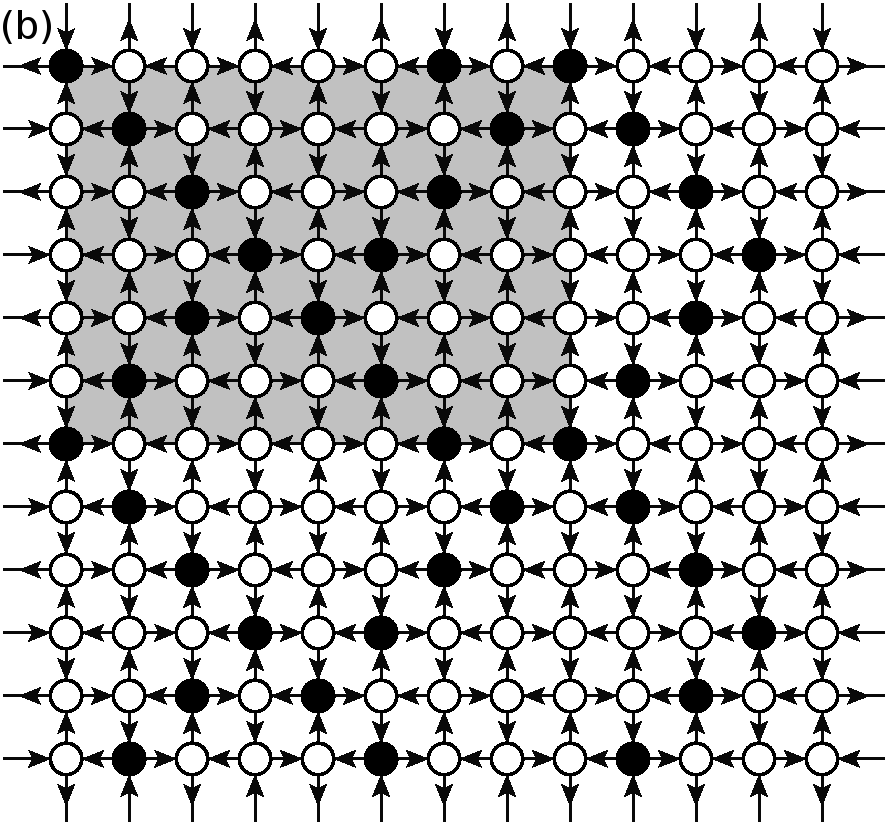}
\caption{The two backgrounds studied on the F-lattice. Unit cells of the periodic
distribution of particles are shown by gray rectangular shades. The filled
circles represent height $0$ and unfilled ones $1$.}
\label{fig:fbg}
\end{center}
\end{figure}

\begin{figure*}
\begin{center}
\includegraphics[width=14cm,angle=0]{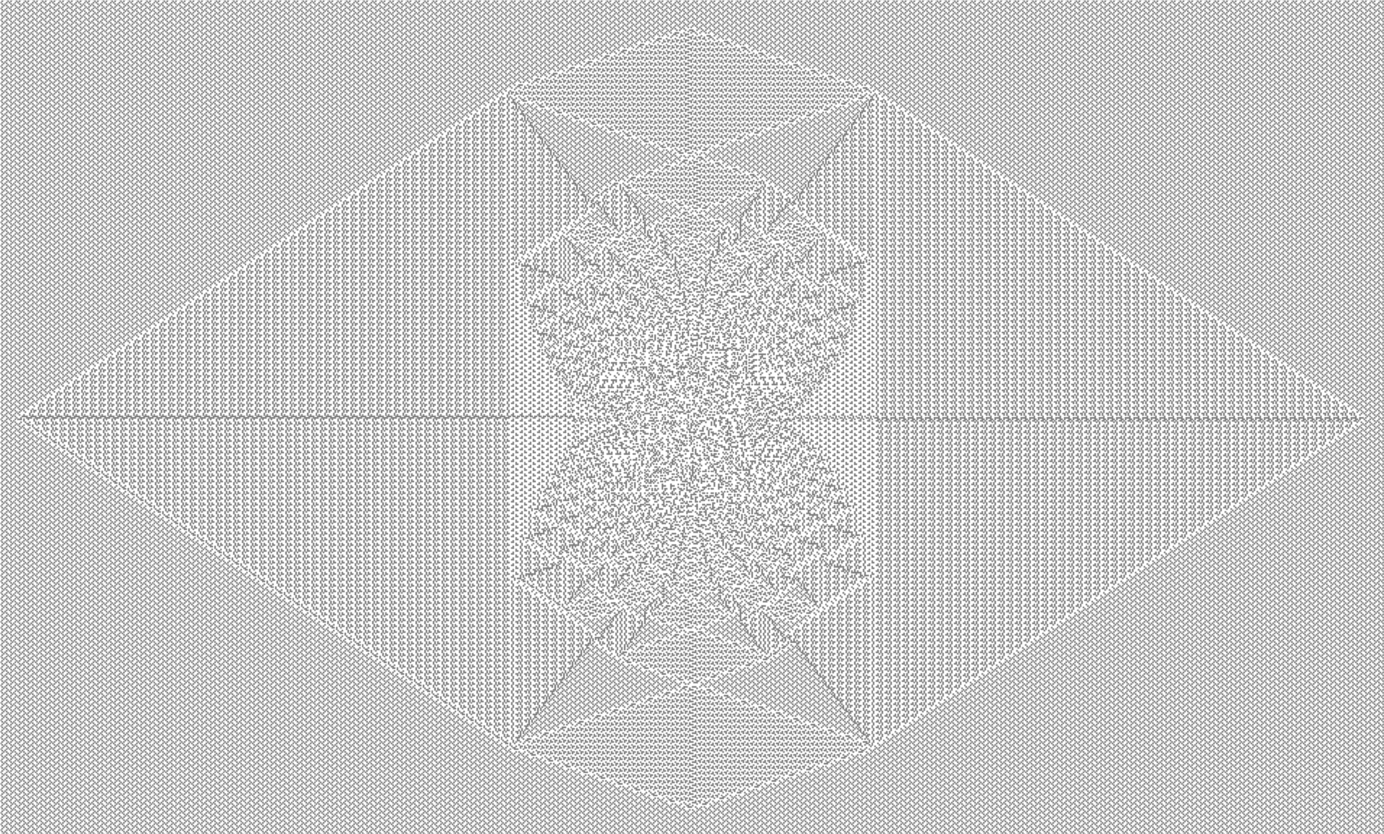}
\caption{The pattern produced on the first background in
Fig. \ref{fig:fbg}, by adding $N=2200$ grains at a single site, and
relaxing the configuration. Color code: White$=1$ and Black$=0$.
Details can be viewed in the electronic
version using zoom in.}
\label{fig:fpic1}
\end{center}
\end{figure*}

\begin{figure*}
\begin{center}
\includegraphics[width=14cm,angle=0]{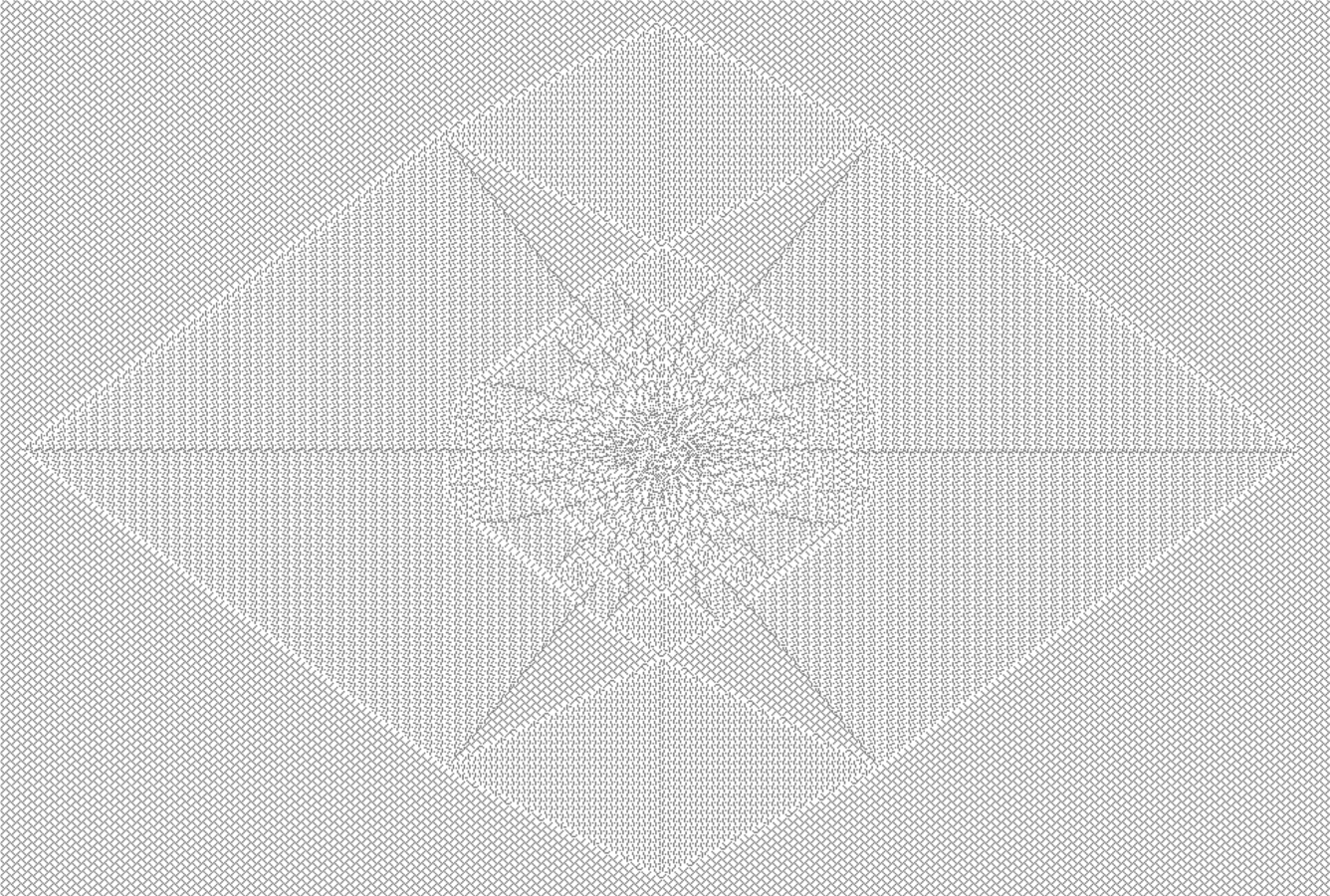}
\caption{The pattern produced on the second background in
Fig. \ref{fig:fbg} by adding $N=600$ grains at a single site, and
relaxing the configuration. Color code: White$=1$ and Black$=0$.
Details can be viewed in the electronic
version using zoom in.}
\label{fig:fpic2}
\end{center}
\end{figure*}

\begin{figure}
\includegraphics[width=9cm,angle=0]{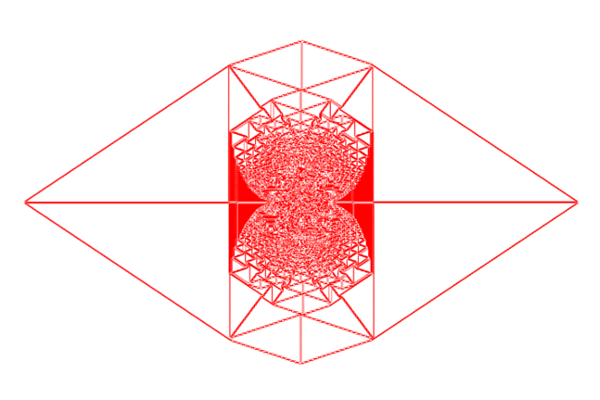}
\caption{The pattern in terms of $Q(r)$, showing the boundaries of 
patches
corresponding to Fig. \ref{fig:fpic1}. Color code: White$=0$ and Red$=$Non-zero.
Details can be viewed in the electronic
version using zoom in.}
\label{fig:flinepic1}
\end{figure}

\begin{figure}
\begin{center}
\includegraphics[width=8cm,angle=0]{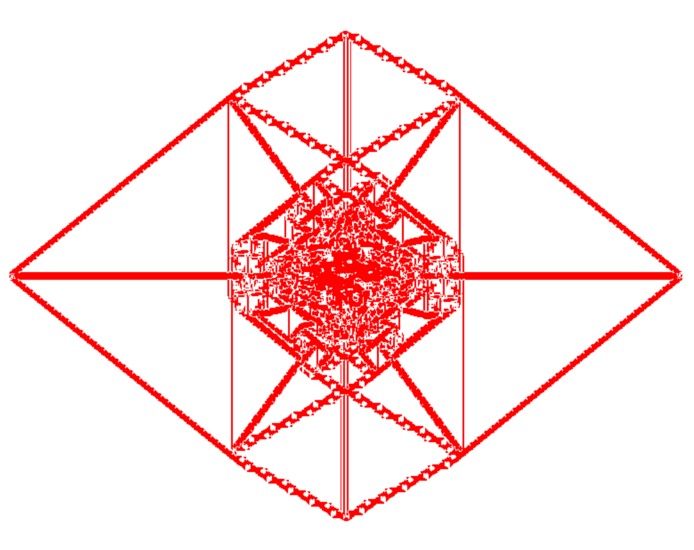}
\caption{The pattern in terms of $Q(r)$, showing the boundaries of patches
corresponding to Fig. \ref{fig:fpic2}. Color code: White$=0$ and Red$=$Non-zero.
Details can be viewed in the electronic
version using zoom in.}
\label{fig:flinepic2}
\end{center}
\end{figure}

\begin{figure}
\includegraphics[width=8cm,angle=0]{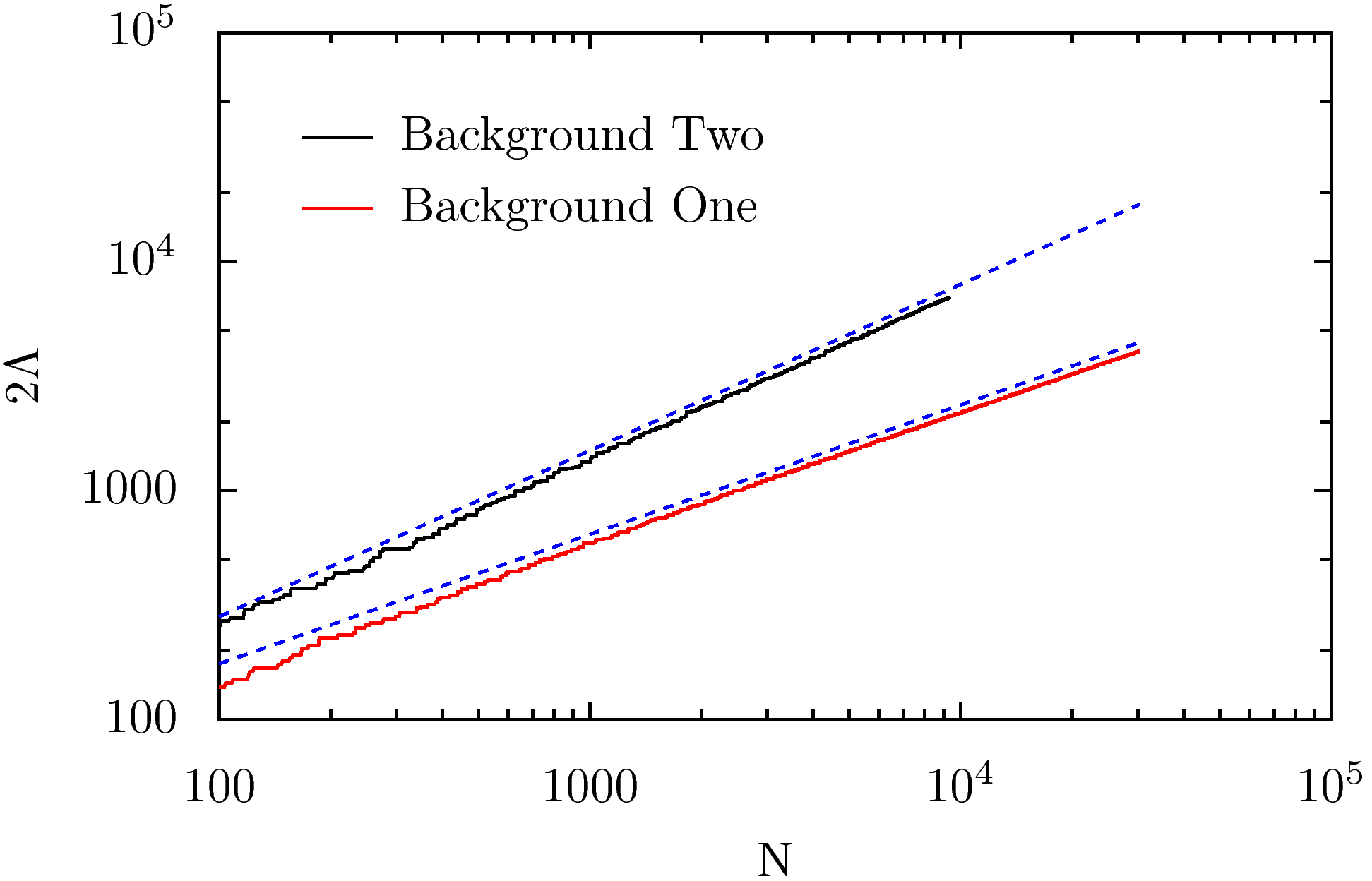}
\caption{The change in diameter as a function of $N$, for the patterns
on the two backgrounds in Fig. \ref{fig:fbg}. The numerical results fit well
with straight lines of slope $0.6$ and $0.725$, for the backgrounds one
and two, respectively.}
\label{fig:flinegrowth}
\end{figure}

On the F-lattice, after some experimentation, we found that  the background pattern 
having the periodicity of the tiling of plane with tilted rectangles,
shown in Fig. \ref{fig:tile_f}, produces patterns with interesting non-compact
growth. We studied rectangles
with aspect ratio $l:(l+1)$, and the rectangles are tilted by
$45^{\circ}$ to the x-axis. Two such periodic backgrounds are shown in
Fig. \ref{fig:fbg}. In
these background patterns, the sites with height zero, are arranged
along the boundaries of tilted rectangles with two
possible orientations, and rest of the sites have
height one. The stable height-patterns generated by adding $N$
particles and relaxing the configuration on these two backgrounds are shown in Fig.
\ref{fig:fpic1} and Fig. \ref{fig:fpic2}, respectively. The growing boundaries of the
patches in the patterns are shown, in
terms of the $Q$ variables, in Fig. \ref{fig:flinepic1} and
\ref{fig:flinepic2}, respectively. Again, we see that the patch boundaries
are straight lines, with rational slopes.
The plot of diameter $2\Lambda$ vs N, for these two patterns are shown in Fig.
\ref{fig:flinegrowth}. We see
that the growth exponent $\alpha $ is approximately $0.6$ for figure
\ref{fig:fpic1} and $0.725$ for figure
\ref{fig:fpic2}. In general, value of the exponent $\alpha$ is in range
$1/2<\alpha <1$, and approaches value $1$ as density
$\rho_{o}$ of the background becomes close to $1$.

There are unresolved areas of apparent solid color in the patterns, taking up a
sizable fraction of the total area, \textit{e.g.},
two large regions of red color on both sides of Fig.
\ref{fig:flinepic1}. In these regions, the pattern appears to be complex,
suggesting either a large number of patch boundaries, or patches of non-zero areal excess charge
density. However, the fractional area of these
regions decreases with larger $N$. Also on comparing patterns
with different $l$, we have seen that the fractional area of such
regions decreases as $l$ increases. A more detailed study of these
patterns seems like an interesting problem for future investigations.
\section{Summary and concluding remarks\label{sec:7}}
In this paper, we have studied two dimensional patterns formed in
Abelian sandpile models by adding particles at one site on an initial
periodic background, where the diameter of the pattern grows as
$N^{\alpha}$, with $\alpha>1/2$. Using some features observed in the
pattern of adjacency of patches as an input, we are able to determine
the exact asymptotic pattern in the  specific case with $\alpha=1$, on
a class $I$ background.

\begin{figure}
\includegraphics[width=8.0cm]{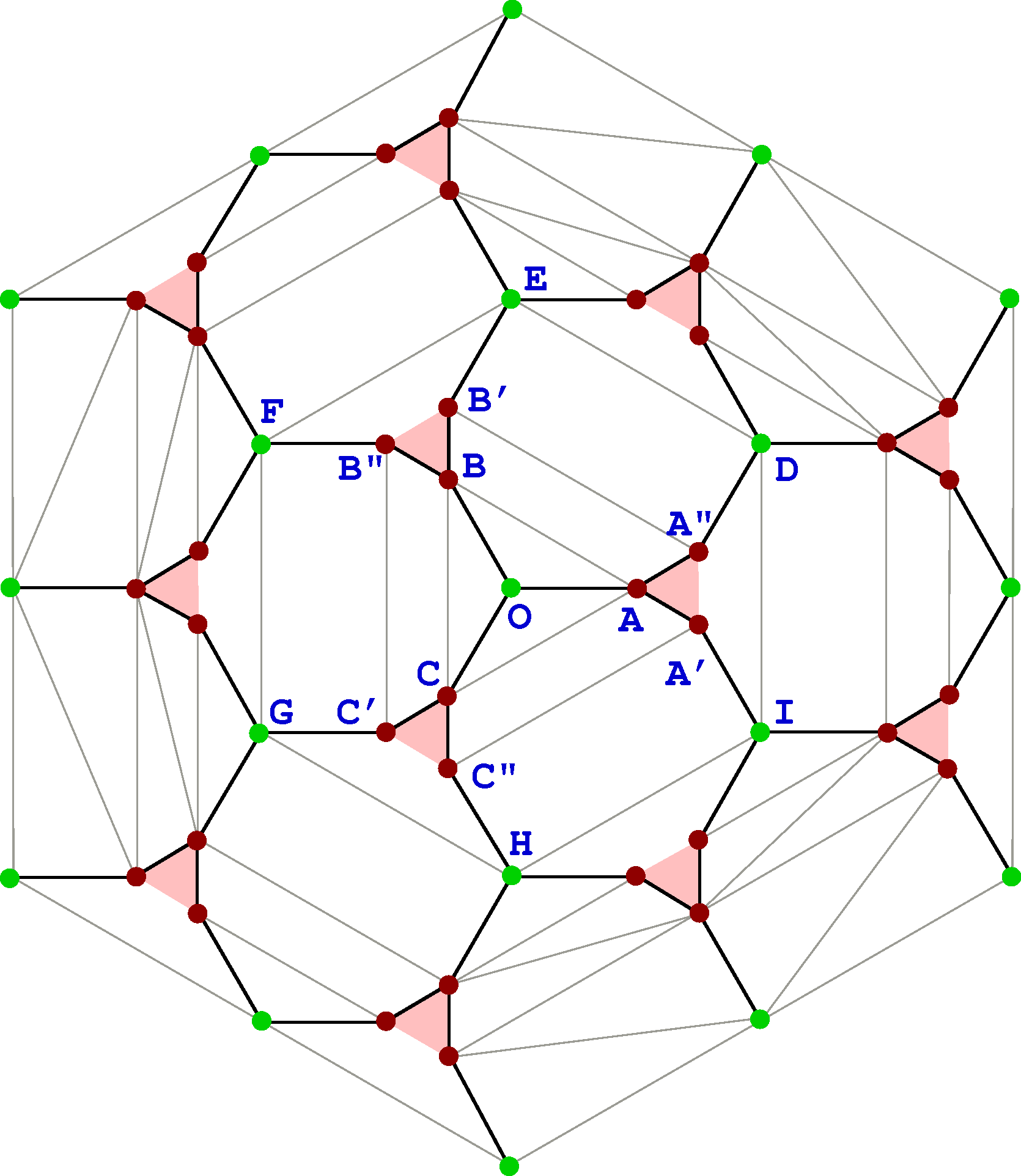}
\caption{The adjacency graph of the patches in the pattern in Fig.
\ref{fig:line}(b). The vertices corresponding to the brownish and greenish
patches in the pattern (Fig.\ref{fig:tri}) are denoted by different colors.}
\label{fig:adjtwo}
\end{figure}

The patterns on class $II$ backgrounds can also be characterized
similarly. As noted earlier, some of the patches split into smaller parts.
By using the $1/\overline{\mathbf{R}}$ transformation, we can again determine the structure of
the adjacency graph. The graph for the pattern in Fig.
\ref{fig:line}(b) is shown in Fig. \ref{fig:adjtwo}. It is a periodic
lattice where half of the vertices of the hexagonal lattice are
replaced by $3$
vertices (colored in brown). The exact D-values for different  patches can be easily
determined. The determination of $f_{m,n}$ for this pattern then
requires the solution of the Laplace's equation on this graph.
It can be shown that a slight alteration of the graph, by
drawing the missing edges in the small triangles shown in pink colors,
does not change the pattern. Then the solution of the Laplace's
equation can be reduced to the solution of a
resistor network on this modified graph. The later can be further
reduced to the resistor network on a
hexagonal lattice, discussed by Atkinson \textit{et.al.}
\cite{atkinson}, using the well-known $Y-\Delta$ transformation.
We omit details of the analysis  here.

An important feature of the non-compact patterns is that,
it can be characterized by a piece-wise
linear function. This characterization is simpler than that of the
patterns with compact growth, where one requires piece-wise quadratic
polynomials. We have shown that there are
infinitely many backgrounds, on which the patterns have non-compact growth.
It would be desirable to determine  the exact value of $\alpha$ for
different backgrounds showing non-compact growth  studied in Sec.
\ref{sec:6}.

\begin{figure}
\includegraphics[width=6cm]{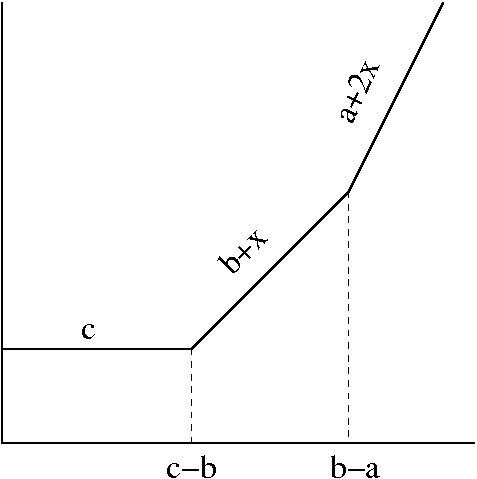}
\caption{Graph corresponding to the tropical function in equation
(\ref{eq:trop})}
\label{fig:trop}
\end{figure}

Another interesting question is a possible connection of this
problem to tropical algebra \cite{tropical}. In tropical mathematics, one defines
operations similar to `addition' and `multiplication' (denoted by $\oplus$ and
$\otimes$ here) by
\begin{eqnarray}
a\oplus b&=&\textrm{max}\left\{ a,b \right\},\nonumber\\
a\otimes b&=&a+b,
\end{eqnarray}
where $a$, $b$ are real parameters.
Familiar properties of addition and multiplication operators, like
commutativity, associativity, existence of identity, distributive
property continues to hold in the new definition. One can then define
polynomials in several variables. The graph of a tropical polynomial
is a piecewise linear function which is also convex. For example, consider the
tropical function
\begin{equation}
f(x)=a\otimes x^{2}\oplus b \otimes x \oplus c.
\label{eq:trop}
\end{equation}
In terms of standard algebra
\begin{equation}
f(x)=max\left\{ a+2x,bx,c\right\}.
\end{equation}
The graph corresponding to this function is shown in Fig. \ref{fig:trop}

\begin{figure}
\includegraphics[width=8cm]{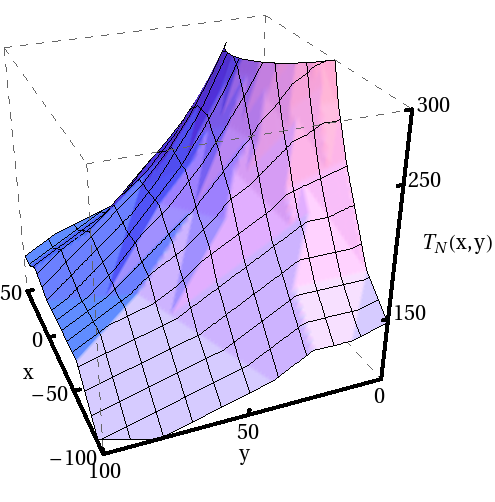}
\caption{(Color online.) Three dimension plot of the integer toppling function
$T_{N}\left( x,y \right)$ for a triangular pattern like in Fig.
\ref{fig:hexpicl1}, but with $N=800$. The plot shows a zoomed-in
section in the region $y\ge0$ and $y+\sqrt{3}x\ge0$.}
\label{surfaceplot}
\end{figure}

We note that for the pattern discussed in section $4$, the potential
function is piece-wise linear. It seems plausible that  tropical polynomials may be useful
to describe this function. In fact, tropical geometry have been
discussed as possibly related to sandpile models \cite{norine,propp2}.
For small values of $N$, our numerical study showed that  $\phi$
is convex, if restricted to one sextant. However, for larger $N$, as
shown in Fig. \ref{surfaceplot}, we see that $\phi$ is not convex even
within one sextant. We conclude that it is not possible to
represent the potential $\phi$ as a simple tropical polynomial. 

\acknowledgements
TS acknowledges the support of Department of Atomic Energy
(DAE), India and Israel Science Foundation
(ISF). DD would like to acknowledge  partial financial support from
the Department of Science and Technology, Government of India, through
a JC Bose Fellowship.

\appendix*
\section{Relation to the theory of discrete analytic functions}
The sandpile patterns we studied are characterized in terms of
discrete analytic functions (DAF) on different discretizations of the
complex plane. For the pattern in Fig. \ref{fig:hexpicl1}, it is the DAF
on a hexagonal lattice, which increases logarithmically at large
distances as in Eq. (\ref{logasymp}).

Studies of DAF started with the work of Kirchhoff on resistor
networks \cite{hu,cserti,Doyle}, and has been studied subsequently by many others
\cite{duffin,mercat}. However, we have not encountered any work on DAF
on many sheeted Riemann surfaces. In the following we present
a way to determine DAF on a square discretization of
Riemann surfaces.

\begin{figure}
\begin{center}
\includegraphics[width=6cm,angle=0]{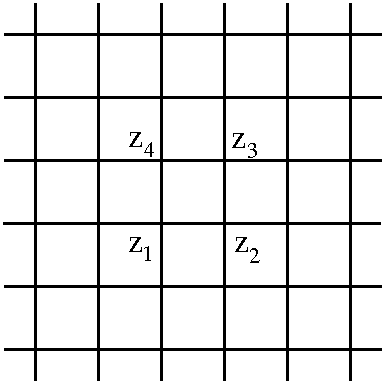}
\caption{\label{fig:square} A square grid on the complex plane.}
\end{center}
\end{figure}

Consider a square grid of points $z=m \epsilon+in\epsilon$, where $m,n$ are
integers and $\epsilon$ is the lattice spacing. Let $f\left(
m\epsilon, n\epsilon \right)$ be a complex function
defined at every site on the grid.
The function $f$ is defined to be discrete analytic
\cite{laszlo} if it satisfies
the discrete Cauchy Riemann condition
\begin{equation}
\frac{f\left(z_{3}\right)-f\left(z_{1}\right)}{z_{3}-z_{1}}=\frac{f\left(z_{4}\right)-f\left(z_{2}\right)}{z_{4}-z_{2}},
\label{eq:dcr}
\end{equation}
at all elementary squares on the grid as shown in Fig. \ref{fig:square}.

In complex analysis, simple examples of analytic functions are
$z^{n}$, and any polynomial of $z^{n}$ is also analytic. For
DAF, it is clear, using the linearity of equation (\ref{eq:dcr}), that
sum of DAF is also discrete analytic. However, not all positive integer powers of
$z$ are discrete analytic. It is easy to check that the functions
$1$, $z$, $z^{2}$, $z^{3}$ are discrete analytic, but $z^{4}$ is not.
We can however construct polynomial functions of $Re(z)$ and
$Im(z)$, that are discrete analytic. Two such examples are
$z^{4}-z\overline{z}\epsilon^{2}$ and $z^{5}-5 z^{2}\overline{z}\epsilon^{2}/2$.

We define a function $F_{n}(z,\epsilon)$ as a homogeneous polynomial in
$z$, $\overline{z}$ and $\epsilon$, of degree $n$, which is a DAF. Using homogeneity, we have
\begin{equation}
F_n(z,\epsilon)=a^n F_n(\frac{z}{a},\frac{\epsilon}{a}),
\end{equation}
and then using $a=\epsilon$, we can express $F_n(z,\epsilon)$ in terms
of $F_n(z,1)$.
This fixes $F_{n}(z,\epsilon)$ up to a multiplicative constant. The
normalization is fixed by requiring that as $\epsilon$ tends to zero,
$F_{n}\left( z, \epsilon \right)\rightarrow z^{n}$.
Then using Cauchy Riemann conditions it is easily seen that
$F_{n}(z, \epsilon)$, for all integers $n\ge0$, has a series expansion in
$\epsilon^{2}$ of the form
\begin{equation}
F_{n}(z,\epsilon)=z^{n}\left[ 1+
\frac{\epsilon^{2}}{z^{2}}g_{1}^{(n)}(\frac{\overline{z}}{z})+\frac{1}{2!}
\frac{\epsilon^{4}}{z^{4}}g_{2}^{(n)}(\frac{\overline{z}}{z})+\cdots
\right],
\end{equation}
where
\begin{eqnarray}
g_{1}^{(n)}(x)&=&-\frac{1}{n-3}\binom{n}{4} x,\\
g_{2}^{(n)}(x)&=&\frac{7!}{(4!)^{2}}\frac{1}{n-6}\binom{n}{7}x^{2}, \\
g_{3}^{(n)}(x)&=&-\frac{10!}{(4!)^{3}}\frac{1}{n-9}\binom{n}{10}x^{3}-\frac{27}{n-7}\binom{n}{8}x,\nonumber\\
\end{eqnarray}
and so on. For an integer $n$, this series will terminate after a
finite number of terms, and all of them can be determined
iteratively.

It is possible to analytically continue the functions for
rational values of $n$. For example,
\begin{equation}
g_{1}^{(n)}(x)=-\frac{\Gamma(n+1)}{4!\Gamma(n-2)}x.
\end{equation}
Then, this analytic continuation of $F_{n}(z,\epsilon)$ provides us the discrete
analytic functions which in the limit $|z|\rightarrow \infty$ grows as
$z^{n}$, for any real positive values of $n$. It is interesting to
note that  the function $D_{m,n}$, used in \cite{myepl} to characterize
the pattern in Fig. \ref{fig:flatone}(b), is equal to
$F_{1/2}(z=m+in,\epsilon=1)$, up to a multiplicative constant. The patterns in the presence of a line
of sinks, or near wedges studied in \cite{myjsp} involve other rational
values. For example, for the pattern near a line sink, one requires
the function $F_{1/3}(z,1)$.
\newpage
\bibliographystyle{unsrt}	
\bibliography{reference}
\end{document}